\documentclass[
  journal=largetwo,
  manuscript=research-article,
  year=2024,
  volume=00,
]{cup-journal}

\usepackage{aas-macros}
\usepackage{amsmath,amssymb}
\usepackage[nopatch]{microtype}
\usepackage{booktabs}
\usepackage{hyperref}

\hypersetup{
    colorlinks,
    citecolor=blue,
    linkcolor=blue,
    urlcolor=blue,
}

\title{Probing magneto-ionic microstructure towards the Vela pulsar using a prototype SKA-Low station}

\author{C.~P.~Lee} 
\affiliation{International Centre for Radio Astronomy Research, Curtin University, Bentley, WA 6102, Australia}
\email[C.~P.~Lee]{christopher.lee@icrar.org}

\author{N.~D.~R.~Bhat} 
\affiliation{International Centre for Radio Astronomy Research, Curtin University, Bentley, WA 6102, Australia}

\author{M.~Sokolowski} 
\affiliation{International Centre for Radio Astronomy Research, Curtin University, Bentley, WA 6102, Australia}

\author{B.~W.~Meyers} 
\affiliation{International Centre for Radio Astronomy Research, Curtin University, Bentley, WA 6102, Australia}

\author{A.~Magro} 
\affiliation{Institute of Space Sciences and Astronomy, University of Malta, Msida, MSD 2080, Malta}


\newcommand{\RMobs}{$\mathrm{RM}_\mathrm{obs}$}
\newcommand{\RMion}{$\mathrm{RM}_\mathrm{ion}$}

\newcommand{\RMism}{$\mathrm{RM}_\mathrm{ISM}$}
\newcommand{\Bpar}{$\langle B_\parallel \rangle$}
\newcommand{\Bparvar}{$\langle B_\parallel \rangle_\mathrm{var}$}
\newcommand{\fctr}{$\nu_\mathrm{ctr}$}
\newcommand{\bw}{$\Delta\nu$}
\newcommand{\rmu}{$\mathrm{rad}\,\mathrm{m}^{-2}$}
\newcommand{\dmu}{$\mathrm{cm}^{-3}\,\mathrm{pc}$}
\newcommand{\rmuyr}{$\mathrm{rad}\,\mathrm{m}^{-2}\,\mathrm{yr}^{-1}$}
\newcommand{\dmuyr}{$\mathrm{cm}^{-3}\,\mathrm{pc}\,\mathrm{yr}^{-1}$}
\newcommand{\dphi}{$\delta\phi$}
\newcommand{\vela}{J0835$-$4510}
\newcommand{\control}{J0630$-$2834}

\begin{document}

\begin{abstract}
The Vela pulsar (\vela) is known to exhibit variations in Faraday rotation and dispersion on multi-decade timescales due to the changing sightline through the surrounding Vela supernova remnant and the Gum Nebula.
Until now, variations in Faraday rotation towards Vela have not been studied on timescales less than around a decade.
We present the results of a high-cadence observing campaign carried out with the Aperture Array Verification System 2 (AAVS2), a prototype SKA-Low station, which received a significant bandwidth upgrade in 2022.
We collected observations of the Vela pulsar and PSR \control{} (a nearby pulsar located outside the Gum Nebula), spanning $\sim$1\,yr and $\sim$0.3\,yr respectively, and searched for linear trends in the rotation measure (RM) as a function of time.
We do not detect any significant trends on this timescale ($\sim$months) for either pulsar, but the constraints could be greatly improved with more accurate ionospheric models.
For the Vela pulsar, the combination of our data and historical data from the published literature have enabled us to model long-term correlated trends in RM and dispersion measure (DM) over the past two decades.
We detect a change in DM of $\sim$0.3\,\dmu{} which corresponds to a change in electron density of $\sim$$10^5\,\mathrm{cm}^{-3}$ on a transverse length scale of $\sim$1--2\,\textsc{au}.
The apparent magnetic field strength in the time-varying region changes from $240^{+30}_{-20}\,\mu\mathrm{G}$ to $-6.2^{+0.7}_{-0.9}\,\mu\mathrm{G}$ over the time span of the data set.
As well as providing an important validation of polarimetry, this work highlights the pulsar monitoring capabilities of SKA-Low stations, and the niche science opportunities they offer for high-precision polarimetry and probing the microstructure of the magneto-ionic interstellar medium.
\end{abstract}

\section{Introduction}\label{sec:intro}
Pulsars are highly-polarised radio sources which have long been used as tools to probe magneto-ionised plasma in the Milky Way and the Local Bubble \citep[e.g.,][]{Manchester1972,Bhat1998}, the solar wind \citep[e.g.,][]{You2012,Howard2016}, and the terrestrial ionosphere \citep[e.g.,][]{Sotomayor-Beltran2013,Porayko2019}.
Propagation through the ionised intervening media imparts frequency-dependent distortions on pulsar signals which can be precisely measured due to the short intrinsic timescales and broadband nature of the signals.
Of particular interest are the cold-plasma dispersion and Faraday rotation effects on the received pulsar signal, characterised by the dispersion measure (DM) and rotation measure (RM) respectively.
Together, these quantities contain information about the thermal electron density and magnetic field strength along the line of sight.
In fact, pulsar observations are exceptionally useful for studying the magneto-ionised interstellar medium (ISM), including the supernova remnants (SNRs) surrounding young pulsars such as the Vela pulsar \citep[e.g.,][]{Hamilton1977,Hamilton1985} and the Crab pulsar \citep[e.g.,][]{Rankin1988}.

There are three standard methods for determining the strength of the magnetic field in the ISM.
Zeeman splitting of spectral lines provides both the strength and direction of the magnetic field; however, it is only observable for relatively strong magnetic fields in H\textsc{i} regions or cold, dense molecular clouds \citep[e.g.,][]{Heiles1976,Heiles1989}.
Alternatively, one can determine the magnetic field strength from the intensity of synchrotron emission, which typically requires making an assumption such as equipartition between the energy density of cosmic ray particles and that of the magnetic field \citep[e.g.,][]{Beck2005,Arbutina2012,Urosevic2018}.
Although the equipartition method is useful when the available data are limited, it is biased towards areas with strong magnetic fields, which leads to estimates which are higher than average \citep{Beck2003}.
Lastly, the magnetic field strength can also be inferred from the Faraday rotation of linearly polarised emission, such as from synchrotron sources or pulsars.
Using pulsar observations, the mean magnetic field strength parallel to the line of sight is simply
\begin{equation}\label{eq:Bpar}
    \langle B_\parallel \rangle \simeq \mathcal{C}^{-1}\frac{\mathrm{RM}}{\mathrm{DM}},
\end{equation}
where $\mathcal{C}^{-1}\approx 1.232\,\mu\mathrm{G}$ and the RM and DM are in their conventional units (\rmu{} and \dmu, respectively).
A positive \Bpar{} indicates that the magnetic field is, on average, pointing towards the observer.
This method is biased towards the warm ionised ISM, and can also be biased if small-scale fluctuations in the magnetic field and the electron density are correlated \citep{Beck2003}.
However, pulsar observations still provide a relatively accurate way to estimate the line-of-sight magnetic field strength in almost any direction, which makes it possible to map the large-scale structure of the Galactic magnetic field \citep[e.g.,][]{Manchester1974,Han2006,Han2018,Sobey2019}.

Stochastic fluctuations in RM and DM due to turbulence in the ISM are expected to be of the order $10^{-5}$--$10^{-4}$\,\rmu{} for observing campaigns of $\sim$1--5\,yr \citep{Porayko2019}.
This is below the sensitivity of current pulsar monitoring campaigns due to the noise floor set by models of the ionospheric RM contribution.
However, it is still possible to study long-term deterministic trends in RM and DM, which can arise for a number of reasons.
Linear trends are often observed due to the transverse motion of the pulsar, which causes the line of sight to probe different regions of the ISM at different times \citep[e.g.,][]{Backer1993,Hobbs2004,You2007,Yan2011,Jones2017,Wahl2022,Keith2024}.
Additionally, periodic variations on a $\sim$1\,yr timescale can arise due to the line of sight probing the solar wind, particularly for pulsars at low solar latitudes \citep[e.g.,][]{Jones2017,Wahl2022,Tiburzi2021}.
Detailed characterisation of these trends is important for reducing red noise in long-term pulsar timing experiments, which is necessary in order to detect the stochastic gravitational wave background \citep[e.g.,][]{Reardon2023,Agazie2023}.

Correlated variations in the measured RM and DM can arise when compact regions of magnetised plasma move through the line of sight.
As demonstrated by \citet{Hamilton1977} and \citet{Hamilton1985}, the mean magnetic field strength in the time-varying region can be estimated from the gradients of the two measures:
\begin{equation}\label{eq:Bfil}
    \langle B_\parallel \rangle_\mathrm{var} \simeq \mathcal{C}^{-1}\left(\frac{\mathrm{dRM}}{\mathrm{d}t}\right)\left(\frac{\mathrm{dDM}}{\mathrm{d}t}\right)^{-1},
\end{equation}
where $\mathcal{C}^{-1}$ is as defined in equation \eqref{eq:Bpar}.
Using observations of the Vela pulsar over $\sim$15\,yr, \citet{Hamilton1985} calculated $\langle B_\parallel \rangle_\mathrm{var}=22\,\mu\mathrm{G}$, which the authors attribute to a magnetised filament in the Vela SNR moving out of the line of sight.
\Bparvar{} has also been estimated for the Crab pulsar \citep[$170\,\mu\mathrm{G}$;][]{Rankin1988} and several other pulsars \citep[see e.g.,][]{vanOmmen1997,Yan2011}.
More recently, \citet{XueThesis} analysed $\sim$50\,yr of historical RM and DM measurements towards the Vela pulsar, and identified three separate monotonic trends in the RM: increasing between 1970--1984, decreasing between 1984--2006, then again increasing between 2006--2019.
Meanwhile, the DM monotonically decreased over this period, with a flattening in the gradient around 1995 as reported by \citet{Petroff2013}.
From equation \eqref{eq:Bfil}, these findings imply that the mean line-of-sight magnetic field through the filament underwent two reversals during this time period.
\citet{XueThesis} suggested that this could be explained by an inhomogeneous magnetic field within the filament.
Given that the Vela pulsar is embedded in a region of turbulent plasma with an apparently inhomogeneous magnetic field, we expect that there may be measurable structure on shorter timescales than what has already been observed, i.e., months to years.

The SKA-Low will be the most sensitive low-frequency telescope ever built, enabling studies of nearly the entire Galactic pulsar population \citep[e.g.,][]{Keane2015}.
In preparation for the construction of the SKA-Low, several prototype `stations' have been deployed at Inyarrimanha Ilgari Bundara, CSIRO's Murchison Radio-astronomy Observatory (MRO), to assist with engineering development.
In this paper, we aim to use one of these stations, the Aperture Array Verification System 2 (AAVS2), to search for temporal changes on timescales of months to years which could arise from magneto-ionic microstructure in the Vela SNR.
Additionally, verification of the prototype station polarimetry will be essential for gaining confidence in polarimetric data obtained with the SKA-Low, and analysis of pulsar observations can be an effective way of achieving this \citep[e.g.,][]{Xue2019}.
The remainder of this paper is organised as follows.
In \S\ref{sec:obs}, we describe the telescope, source selection, observations, and data reduction.
In \S\ref{sec:analysis}, we describe the methods used to estimate the RM and DM.
In \S\ref{sec:results}, we present the results, including polarimetric pulse profiles and the RM and DM measurements.
We use these results to validate the station polarimetry and test ionospheric models.
In \S\ref{sec:discussion}, we compare the results with historical data and discuss implications for future low-frequency monitoring science.
Finally, in \S\ref{sec:conclusions}, we summarise and present our conclusions.

\section{Observations and data reduction}\label{sec:obs}
All observations were collected with the AAVS2 \citep{vanEs2020,Macario2022}, a prototype SKA-Low station consisting of 256 dual polarised log-parabolic antennas pseudo-randomly distributed over a circular ground plane with a diameter of $\sim$42\,m and a maximum baseline of $\sim$38\,m.
The AAVS2 operates in the frequency range 50--350\,MHz and records $\sim$925.926\,kHz wide coarse channels separated by $\sim$781.25\,kHz in order to avoid the drop in sensitivity towards the edges of the bandpass response.
The initial AAVS2 system was only capable of recording data from a single coarse channel at a time, but nevertheless was used to detect 20 pulsars at various frequencies \citep{Lee2022}.
In early 2022, the AAVS2 received an upgrade which enabled it to record data from multiple contiguous coarse channels.
Although in principle this upgrade enables the simultaneous recording of an arbitrary number of channels, the practical limit is set by the maximum possible data recording rate.
In this work, we recorded contiguous frequency bands with 16 and 32 channels, corresponding to instantaneous bandwidths of 12.5\,MHz and 25\,MHz, respectively.
In Table \ref{tab:observations}, we summarise the frequency setups used and their corresponding effective resolutions in Faraday space as defined in equation \eqref{eq:phires}.

\begin{table}[t]
    \centering
    \caption{AAVS2 frequency bands used in this work. The columns (from left to right) are: the centre frequency (\fctr), the bandwidth (\bw), the span in $\lambda^2$ ($\Delta\lambda^2$), the Faraday depth resolution (\dphi; see equation \ref{eq:phires}), the number of observations ($N_\mathrm{obs}$), and the target pulsar.}
    \label{tab:observations}
    \begin{tabular}{cccccc}
        \hline
        \fctr & \bw & $\Delta\lambda^2$ & \dphi & $N_\mathrm{obs}$ & Target \\
        $\text{[}$MHz] & [MHz] & [$\text{m}^2$] & [$\text{rad}\,\text{m}^2$] & & \\
        \hline
        105.859375 & 12.5 & 1.91 & 1.99 & 30 & \control \\
        205.859375 & 12.5 & 0.26 & 14.73 & 22 & \vela \\
        212.109375 & 25.0 & 0.47 & 8.01 & 2 & \vela \\
        326.171875 & 12.5 & 0.06 & 58.64 & 27 & \vela \\
        \hline
    \end{tabular}
\end{table}

We collected multi-channel observations of PSR \vela{} (the Vela pulsar) between 2022-02 and 2022-11 to assist with testing and verification of the upgraded recording system.
Using these initial observations, we selected the frequency bands to use for further monitoring.
The precision of RM measurements increases at lower frequencies due to the increasing number of rotations of the electric vector over the observing bandwidth.
Therefore, we obtain the best precision when observing at the lowest possible frequencies.
The Vela pulsar is heavily scattered at low frequencies, becoming undetectable in beamformed observations below $\sim$160\,MHz \citep[e.g.,][]{Kirsten2019}.
Since heavy scattering depolarises and degrades the signal, we selected a frequency range between 199.6\,MHz and either 212.1 or 224.6\,MHz (for 16 and 32-channel recording bands, respectively).
We also selected a higher frequency band between 319.9--332.4\,MHz to use for additional validation of the AAVS2 polarimetry, as well as DM measurements.
An observing campaign was carried out on the Vela pulsar with a cadence of $\sim$1\,day--3\,weeks (depending on the telescope availability) between 2022-11-21 and 2023-05-23.
In this paper, we use the observations from this monitoring campaign, as well as some additional observations in the same frequency ranges which were initially recorded for engineering verification purposes.
The observations were typically 15\,min in length and collected at transit ($\sim$18\,deg from zenith).

In addition to the Vela pulsar, we collected observations of PSR \control{} in the frequency range 99.6--112.1\,MHz.
This pulsar has a sky position outside of the Gum Nebula and is strongly polarised with no significant scatter broadening in this frequency band.
It also has a high fractional linear polarisation and transits near zenith, making it a good control source for polarimetric verification \citep[e.g.,][]{Xue2019}.
Five observations were collected at source transit on various dates between 2022-08-14 and 2022-12-09.
The remaining observations were collected in sets of nine, where four observations were collected before and after transit in addition to an observation at transit.
Three observations were discarded due to problems with the recording system, leaving 24 observations at various angles from zenith.

The beamformed voltages were coherently dedispersed and folded using \texttt{DSPSR} \citep{vanStraten2011}, and subsequent processing was performed using utilities from \texttt{PSRCHIVE} \citep{Hotan2004,vanStraten2012}.
The pulsar archives were processed with 256 phase bins (equivalent to a time resolution of 348.93\,$\mu$s for Vela) and a frequency resolution of 3.617\,kHz.
The oversampled bandwidth was removed and the coarse channels were combined using \texttt{psradd}.
The topocentric folding period and DM were then updated using \texttt{pdmp} and \texttt{pam}, and radio frequency interference (RFI) was excised using \texttt{paz}.
The Faraday rotation imparted on each observation was removed using \texttt{pam}.

\section{Data analysis}\label{sec:analysis}
\subsection{RM-synthesis}

\begin{figure*}[t]
    \centering
    \includegraphics[width=0.495\linewidth]{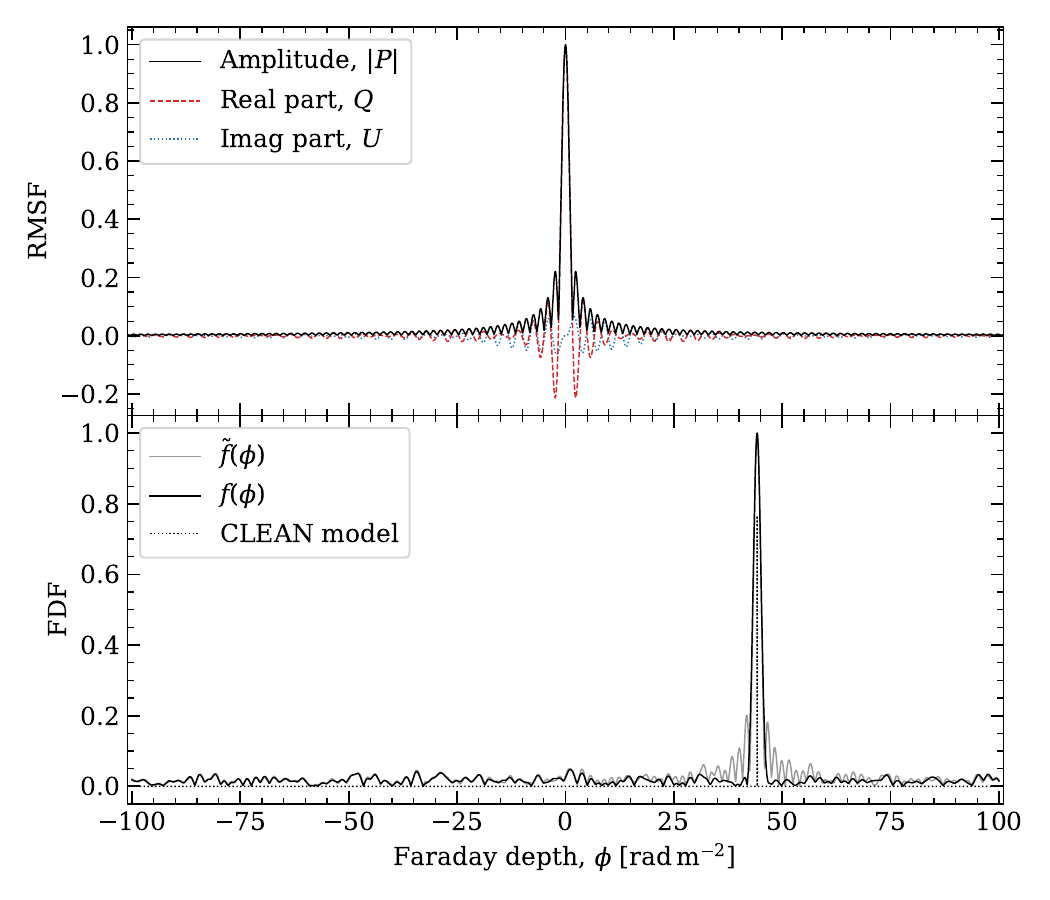}
    \includegraphics[width=0.495\linewidth]{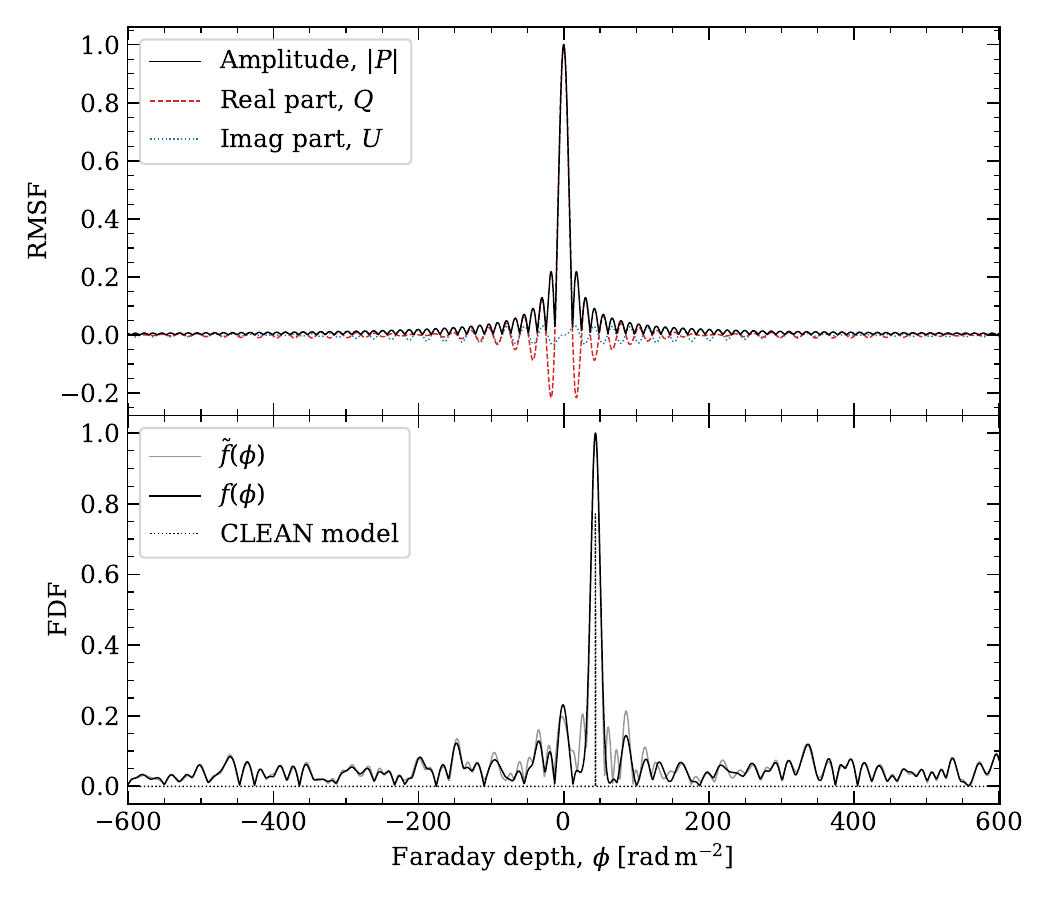}    
    \caption{Example RM-synthesis results for observations of \control{} centred at 105.9\,MHz (left) and \vela{} centred at 205.9\,MHz (right). The bandwidth of both observations is 12.5\,MHz. Each subfigure shows the following. Top: The RMSF, including the real ($Q$) and imaginary ($U$) parts. Bottom: The Faraday spectrum before [$\tilde{f}(\phi)$] and after [$f(\phi)$] deconvolution of the RMSF, with the RM-CLEAN model shown. The highest peak in the FDF corresponds to the measured RM, and the smaller peak at $\phi\sim0\,$\rmu{} for \vela{} is caused by instrumental polarisation.}
    \label{fig:RMsyn-eg1}
\end{figure*}

The polarisation state of an electromagnetic wave can be described by the four Stokes parameters $(I, Q, U, V)$, where $I$ is the total intensity, $Q$ and $U$ are the components of linear polarisation, and $V$ is the circular polarisation.
The linear polarisation is described by a vector in the complex plane
\begin{equation}
    P = Q + iU = pI e^{2i\psi},
\end{equation}
where $p$ is the fractional degree of linear polarisation and $\psi$ is the position angle of linear polarisation.
Propagation through magnetised plasma rotates the polarisation vector as a function of the observational wavelength ($\lambda$) squared.
When the polarised signal originates from a single source along the line of sight, the rotation in $\lambda^2$ is linearly proportional to the RM, such that
\begin{equation}\label{eq:PA}
    \psi = \psi_0 + \mathrm{RM}\lambda^2,
\end{equation}
where $\psi_0$ is the intrinsic position angle and
\begin{equation}\label{eq:rm}
    \mathrm{RM} = \mathcal{C}\int_\mathrm{source}^\mathrm{observer} n_e(l) B_\parallel(l)\,\mathrm{d}l,
\end{equation}
where $\mathcal{C}$ is as defined in equation \eqref{eq:Bpar}, $n_e$ is the electron density ($\mathrm{cm}^{-3}$), $B_\parallel$ is the interstellar magnetic field component parallel to the line of sight ($\mu\mathrm{G}$), and $\mathrm{d}l$ is an infinitesimal path length (pc).
The traditional approach to finding the RM is to fit a linear model to $\psi$ as a function of $\lambda^2$ and measure the gradient.
However, in recent years it has become more common to use the method of RM-synthesis \citep{Burn1966,Brentjens2005}, which uses a Fourier transform of the polarisation vector to determine the RM.
This approach is more reliable for low signal-to-noise observations, and allows for the separation of the astrophysical signal from weakly chromatic instrumental effects (such as leakages between polarisations).

Following \citet{Brentjens2005}, we introduce a weighting function $W(\lambda^2)$ defined as non-zero at every $\lambda^2$ where measurements are made and zero elsewhere.
We then express the observed complex polarisation vector as $\tilde{P}_i = w_iP(\lambda_i^2)$, where $\lambda_i$ is the centre wavelength of channel $i$ and $w_i=W(\lambda_i^2)$.
The Faraday dispersion function (FDF; also referred to as the Faraday spectrum) is the Fourier transform of the observed complex polarisation vector,
\begin{equation}\label{eq:rFDF}
    \tilde{f}(\phi) \simeq K\sum_{i-1}^N \frac{\tilde{P}_i}{s_i} e^{-2i\phi(\lambda_i^2-\lambda_0^2)},
\end{equation}
where $\lambda_0$ is the reference wavelength \citep[as shown by][this is ideally the weighted average of $\lambda_i^2$]{Brentjens2005}, $\phi$ is the Faraday depth (a generalisation of RM defined at all points along the line of sight), $K$ is the reciprocal of the sum of all weights, and $s_i=s(\lambda_i^2)$ where $s(\lambda^2)=I(\lambda^2)/I(\lambda_0^2)$ is a function describing the spectral dependence of the total intensity.
The FDF represents the polarised flux density at every Faraday depth; it contains a single peak at the RM as long as the polarisation originates from a single source with no internal Faraday rotation (i.e., when the Faraday rotation is proportional to $\lambda^2$ as in equation \ref{eq:PA}).
In this paper, we use RM to refer to the measured Faraday depth at the peak of the FDF.

The synthesised FDF in equation \eqref{eq:rFDF} is the convolution of the `true' FDF with the RM spread function (RMSF),
\begin{equation}
    R(\phi) \simeq K\sum_{i-1}^N w_i e^{-2i\phi(\lambda_i^2-\lambda_0^2)}.
\end{equation}
In order to obtain the best signal-to-noise ratio, the RMSF must be deconvolved from the synthesised FDF.

The resolution in Faraday space is determined by the full width at half maximum (FWHM) of the RMSF,
\begin{equation}\label{eq:phires}
    \delta\phi \left[\mathrm{rad}\,\mathrm{m}^{-2}\right] = \frac{3.8}{\lambda_\mathrm{max}^2 - \lambda_\mathrm{min}^2},
\end{equation}
where $\lambda_\mathrm{min}$ and $\lambda_\mathrm{max}$ are the shortest and longest wavelengths in metres, and we use the updated proportionality constant from \citet{Schnitzeler2009}.
We then estimate the uncertainty in $\phi$ as 
\begin{equation}
    \sigma_\phi = \frac{\delta\phi}{2.355\times \mathrm{S}/\mathrm{N}_F},
\end{equation}
where $\mathrm{S}/\mathrm{N}_F$ is the polarised signal-to-noise ratio in the FDF and 2.355 is the conversion between the FWHM and the standard deviation of a Gaussian.

Our analysis was performed using a Python implementation of RM-synthesis and the RM-CLEAN deconvolution algorithm made available as a public repository\footnote{\url{https://github.com/gheald/RMtoolkit}} by \citet{Heald2009}.
To measure the RM, the observation archives were first downsampled to 32 phase bins to increase the signal-to-noise of the Stokes $(I,Q,U)$ spectra per bin.
The bin with the highest total intensity was then extracted using \texttt{pdv} from \texttt{PSRCHIVE}.
The Stokes $Q$ and $U$ samples were normalised by the best-fit power-law model to the Stokes $I$ samples as a function of frequency, then combined to form samples of the polarisation vector, $P_i=w_i[Q(\lambda_i)+iU(\lambda_i)]$, using a uniform weighting scheme.
For the observations of \control, the FDF was computed in the Faraday depth range $-100\leq\phi\leq 100$\,\rmu{} with a step size of 0.001\,\rmu; whilst for Vela, we used the range $-600\leq\phi\leq 600$\,\rmu{} and a step size of 0.01\,\rmu.
These parameter choices ensure that the FDF is oversampled (to precisely locate the peak) and enough off-peak noise is captured for calculating $\mathrm{S}/\mathrm{N}_F$.
The RMSF was deconvolved from the FDF using an RM-CLEAN component cutoff of $\mathrm{S}/\mathrm{N}_F=2$.
The RM was then estimated by fitting a parabola to the peak of the FDF and solving for the Faraday depth of the vertex.
The $\mathrm{S}/\mathrm{N}_F$ was calculated as the ratio of the peak in the FDF and the standard deviation of the RM-CLEAN residuals.

Figure \ref{fig:RMsyn-eg1} shows example FDFs and RMSFs for \control{} and \vela.
For both pulsars, the FDF shows a clear peak at the RM of the pulsar.
The \vela{} FDF shows a small peak at $\phi\sim 0$\,\rmu{} caused by instrumental polarisation, whilst the \control{} FDF shows no significant instrumental peak.
Due to the pristine RFI conditions at the MRO, the observations required very minimal data flagging, particularly in the $\sim$200\,MHz frequency bands.
As a result, the theoretical RMSFs are all similar in shape, with a main lobe flanked by increasingly weaker sidelobes.
The resolution in Faraday space is significantly smaller for \control, but in both cases the resolution is high enough to neglect any instrumental polarisation.
For the frequency band at 326.17\,MHz, the $\phi$ resolution is not sufficient to resolve the RM peak from the instrumental peak, so reliable RM measurements could not be obtained for these observations.
Nevertheless, we were able to measure an apparent RM in order to remove the majority of the Faraday rotation from the data.

\subsection{Ionospheric RM subtraction}
The observed RM (\RMobs) can be treated as a sum of the contributions from different Faraday rotating media along the line of sight.
The main contributions come from the ionised ISM (\RMism), the interplanetary medium, and the terrestrial ionosphere (\RMion).
In practice, the interplanetary plasma (from solar wind and coronal mass ejections) only needs to be considered for pulsars near the ecliptic \citep[e.g.,][]{You2012}, which is not the case for the pulsars studied in this work.
We therefore assume the following:
\begin{equation}
    \mathrm{RM}_\mathrm{obs} = \mathrm{RM}_\mathrm{ISM} + \mathrm{RM}_\mathrm{ion}.
\end{equation}
In order to estimate $\mathrm{RM}_\mathrm{ISM}$, the ionospheric contribution must be modelled and subtracted from \RMobs.
An ideal ionospheric model must account for variability on a wide range of timescales, including solar flares (on timescales of minutes), diurnal variations (1\,day cycle), and the 11\,yr solar cycle \citep[e.g.,][]{Sotomayor-Beltran2013,Lam2016}.
There are several publicly available software repositories which can be used to estimate \RMion.
In this paper, we use the Python package \texttt{RMextract}\footnote{\url{https://github.com/lofar-astron/RMextract}} \citep{RMextract}.
We provide a brief summary of the methods implemented in this software below, but further details can be found in \citet{Porayko2023}.

For simplicity, it is common practice to approximate the ionosphere as an infinitesimally thin shell of plasma with an effective altitude of between 350--650\,km \citep[known as the single layer model or SLM; e.g.,][]{Sotomayor-Beltran2013}.
The value of $\mathrm{RM}_\mathrm{ion}$ at the intersection of the line of sight and the ionospheric shell, known as the ionospheric pierce point (IPP), is estimated as
\begin{equation}\label{eq:slm}
    \mathrm{RM}_\mathrm{ion} = \eta\,\mathrm{TEC}_\mathrm{LOS}B_\mathrm{LOS},
\end{equation}
where $\eta=2.630\times 10^{-17}\,\mathrm{G}^{-1}$, $\mathrm{TEC}_\mathrm{LOS}$ is the slant total electron content at the geographical location of the IPP ($\mathrm{m}^{-2}$) and $B_\mathrm{LOS}$ is the geomagnetic field strength at the IPP (G).
The total electron content (TEC) can be obtained from global ionosphere maps (GIMs) provided by the International GNSS Service \citep[IGS;][]{HP2009} via the NASA Archive of Space Geodesy Data\footnote{\url{https://cddis.nasa.gov/archive/gps/products/ionex/}}.
There are several GIMs provided by different analysis centres, each sharing data from the same network of IGS ground stations but using different modelling and interpolation methods.
\citet{Porayko2019} performed a detailed comparison of several GIMs for the site of the Low-Frequency Array (LOFAR), finding that the best results came from UQRG \citep[from the Technical University of Catalonia;][]{Orus2005} and JPLG (from the Jet Propulsion Laboratory).
Since the accuracy depends on the geographical coverage of GNSS ground stations, additional verification is required at the MRO.
Although a similarly detailed comparison is beyond the scope of this paper, we provide a simple comparison of UQRG and JPLG on SKA-Low station data in \S\ref{sec:verification}.
The geomagnetic field ($B_\mathrm{LOS}$) was calculated using the World Magnetic Model \citep{WMM}.
Other models are available, however \citet{Porayko2019} found that the difference between models is $\sim 0.1\%$, which is negligible compared to the uncertainties in the ionospheric TEC.
\texttt{RMextract} spatially interpolates the vertical TEC maps to the IPP using a simple four-point formula \citep[e.g.,][]{IONEX}.
The maps are then computed for the observation epoch using a linear interpolation in time.

As discussed by \citet{Porayko2023}, the full electron density profile of the atmosphere includes the plasmasphere, which extends from the top of the ionosphere up to $\sim$20,000\,km in altitude.
Since the RM is an integrated quantity, the magnetic field strength gradient over this altitude range can cause an overestimation in the RM determined using the single-layer approximation.
To account for this, \texttt{RMextract} includes a second method which numerically integrates equation \eqref{eq:slm} over the electron density and magnetic field strength profiles, then normalises the integral by the vertical TEC values from the GNSS GIMs.
The ionospheric density profile is obtained from the International Reference Ionosphere (IRI) model and its extension to plasmaspheric altitudes (IRI-Plas).

Following \citet{Sotomayor-Beltran2013}, we used the root-mean-square (RMS) vertical TEC maps to estimate the $1\sigma$ uncertainty for each \RMion{} measurement.
Inspection of the estimated uncertainties revealed several outliers ranging from $\sim$0.07--1.2\,\rmu, whereas the vast majority of points were distributed close to the median value of $\sim$0.2\,\rmu.
We therefore opted to use an uncertainty of 0.2\,\rmu{} for all measurements.

\subsection{Determination of DM}
To determine the DM, the data was averaged into 8 frequency subbands and a template profile was created by co-adding all of the available observations for the given frequency band.
We obtained pulse arrival times for each subband using the Fourier phase gradient algorithm with Markov chain Monte Carlo error estimation, as implemented in the \texttt{PSRCHIVE} routine \texttt{pat}.
We then fit the arrival times to measure the DM using \texttt{Tempo2} \citep{Hobbs2006}.

\section{Results}\label{sec:results}
\subsection{Polarimetric pulse profiles}
\begin{figure*}[t]
    \centering
    \includegraphics[width=\linewidth]{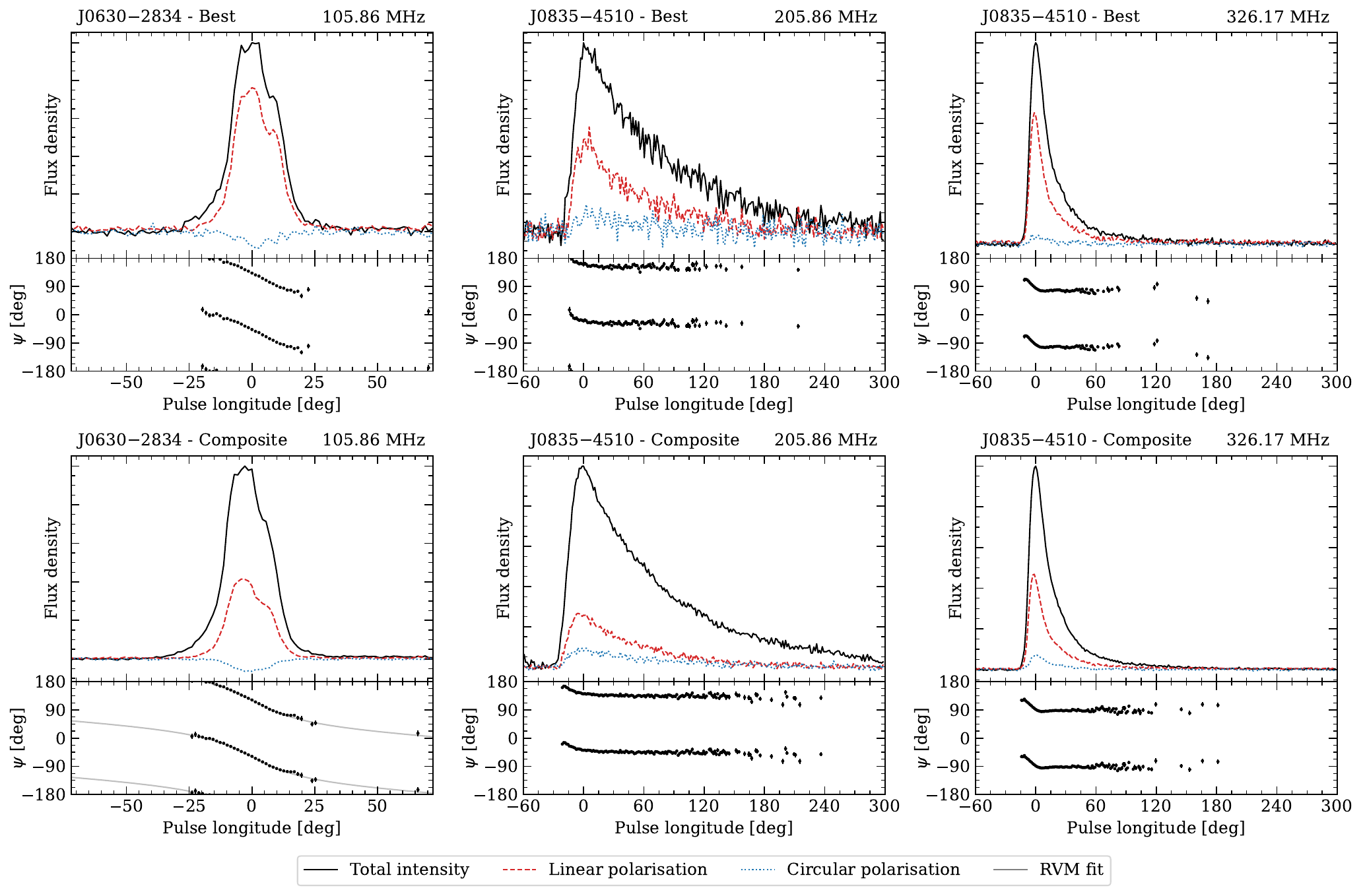}
    \caption{Full-Stokes integrated pulse profiles for \control{} at 105.86\,MHz (left); and \vela{} at 205.86\,MHz (middle) and 326.17\,MHz (right). All profiles are processed with 256 phase bins. The top row shows the best single observation based on the linearly polarised signal-to-noise ratio. The bottom row shows composite profiles of all observations in the given frequency band. For each subplot, the top panel shows the pulse profile in total intensity (Stokes $I$; black line), linear polarisation (Stokes $\sqrt{Q^2+U^2}$; red dashed line), and circular polarisation (Stokes $V$; blue dotted line). The lower panel shows the position angle for bins with $>3\sigma$ in linear polarisation. For the composite profile of \control, the best-fit rotating vector model (RVM) is shown (grey line).}
    \label{fig:polprofs}
\end{figure*}

\begin{figure}[t]
    \centering
    \includegraphics[width=\linewidth]{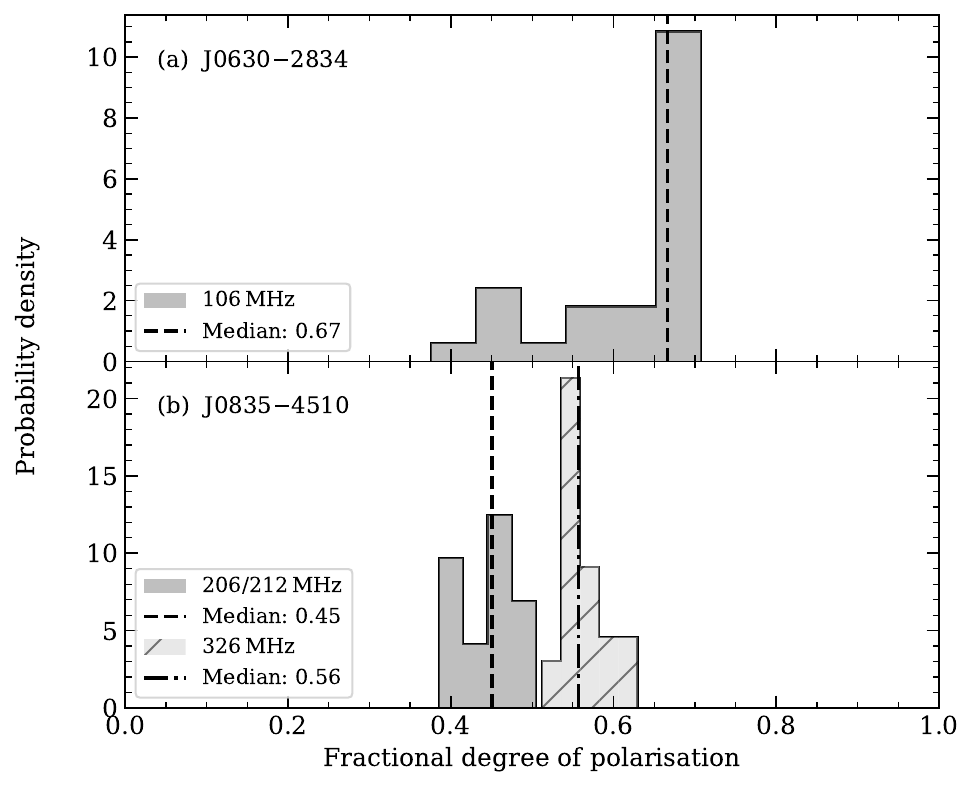}
    \caption{Histograms of the fractional degree of linear polarisation measured from AAVS2 detections. (a) Result for \control{} at 105.86\,MHz. (b) Result for \vela{} at 205.86/212.11\,MHz (dark grey) and 326.17\,MHz (light grey, hatched). The median values are indicated by dashed and dot-dashed lines.}
    \label{fig:fracpols}
\end{figure}

Full-Stokes integrated pulse profiles can provide useful insights into the accuracy and reproducibility of polarimetry.
This is particularly important for the AAVS2, a prototype aperture array with novel technologies that are yet to be fully validated.
In Figure \ref{fig:polprofs}, we present polarimetric pulse profiles for both of the target pulsars.
The top row shows the best detections for each frequency band, selected based on the linearly polarised signal-to-noise.
The bottom row shows composite profiles created by phase-aligning and combining all of the integrated pulse profiles with uniform weights using \texttt{psradd}.
The Faraday rotation was removed for each observation before the profiles were combined.
For \control, the best individual detection is consistent with detections from the Giant Metrewave Radio Telescope (GMRT) at 243\,MHz \citep{Johnston2008} and the Murchison Widefield Array \citep[MWA;][]{Tingay2013,Wayth2018} between 140 and 200\,MHz \citep{XueThesis}.
At 326\,MHz, the best individual profile of \vela{} is consistent with the detection from \citet{Hamilton1977b} made using Murriyang (CSIRO's Parkes 64-m radio telescope).
At 206\,MHz, the best AAVS2 detection is consistent with the full-Stokes profile from the MWA at the same frequency \citep{XueThesis}.
The composite profiles display a significantly lower degree of polarisation than the individual observations.
This is indicative of some residual error in the polarimetry, which will require further investigation to resolve.

The degree of linear polarisation is expected to be intrinsically stable, which makes it possible to quantify the stability of the polarimetry between observations.
For each profile, we measured the fractional degree of linear polarisation ($p$) and present the results in Figure \ref{fig:fracpols}.
For \control, half of the observations have a fractional polarisation between 0.67 and 0.71, which is similar to what is reported by \citet{Johnston2008} and \citet{XueThesis}.
However, the other half of the observations show a large spread of fractional polarisations, indicating some inconsistency in the quality of the uncalibrated polarimetry.
We did not find any significant correlation between the fractional polarisation and the zenith angle, which suggests that the systematic depolarisation is not direction dependent.
The fractional polarisations of \vela{} are more consistent, with some variance that could be caused by the changing pulse broadening time between observing epochs \citep[e.g.,][]{XueThesis}.
The range of fractional polarisations are consistent with measurements from \citet{Hamilton1977b} and \citet{XueThesis}, with minor discrepancies which could be due to frequency-dependent depolarisation from stochastic fluctuations in the ISM \citep{Burn1966}.

For \control, we fit the rotating vector model \citep[RVM;][]{Radhakrishnan1969} using \texttt{PSRSALSA} \citep{Rookyard2015,Weltevrede2016}.
We used the routine $\texttt{ppolFit}$, which determines the allowed values of the magnetic inclination angle ($\alpha$) and the impact parameter ($\beta$) by performing a grid search in the $\alpha$--$\beta$ parameter space and measuring the reduced $\chi^2$ for each trial.
The 68\% confidence interval of $\alpha$ and $\beta$ was determined from the points at which the reduced $\chi^2$ was equal to twice the global minimum.
Using the AAVS2 composite profile at 105.86\,MHz, we measure $0\leq\alpha\leq 87.5\,\mathrm{deg}$ and $-13.6\leq \beta\leq 0\,\mathrm{deg}$.
For comparison, we performed the same analysis on a pulse profile published by \citet{Johnston2018} at 1.4\,GHz.
In this case, we find $0\leq\alpha\leq 92.8\,\mathrm{deg}$ and $-14.1\leq \beta\leq 0\,\mathrm{deg}$.
The AAVS2 measurements are therefore consistent with \citet{Johnston2018}, and place marginally stronger constraints on $\alpha$ and $\beta$.
In the lower left subplot of Figure \ref{fig:polprofs}, we show the RVM fit with the minimum reduced $\chi^2$; that is, $\alpha=0.52\,\mathrm{deg}$ and $\beta=-0.12\,\mathrm{deg}$.

\subsection{Validation of ionospheric models}\label{sec:verification}
Due to the large fractional bandwidth, the AAVS2 is capable of determining pulsar RMs with exceptional precision.
For our observations of \control{} centred at 106\,MHz, 95\% of the RM uncertainties are between $\sim$0.01--0.04\,\rmu.
For comparison, the 95\% confidence lower bound of the distribution of RM uncertainties reported in the ATNF pulsar catalogue (v1.71) is $\sim$0.06\,\rmu{} \citep{Manchester2005}.
It is therefore clear that the AAVS2 can determine RM to a precision significantly better than telescopes operating at higher frequencies.
This feature makes the AAVS2 well suited for monitoring small variations in ionospheric RM on short timescales ($\sim$hours), which is useful for testing the accuracy of ionospheric models.

\begin{figure}[t]
    \centering
    \includegraphics[width=\linewidth]{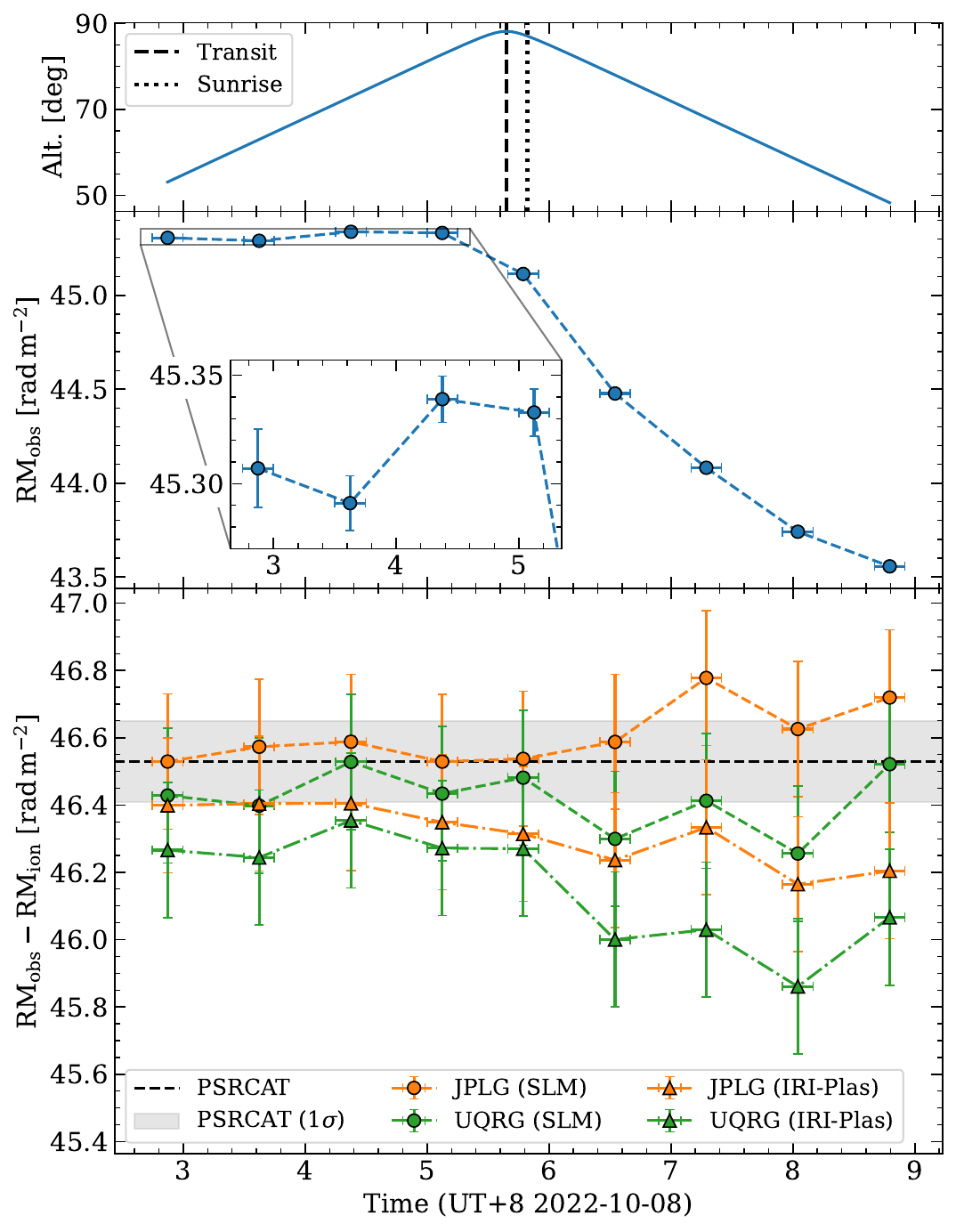}
    \caption{Temporal RM variations towards \control{} over a $\sim$6\,h period on UT+8 2022-10-08. The top plot shows the source altitude over time, transiting just before sunrise (the time when the Sun reaches an altitude of 0\,deg). The middle plot shows the observed RM; the inset shows the small RM variations detected before sunrise. The bottom plot shows the ionosphere-subtracted RMs for JPLG (orange) and UQRG (green) TEC maps with the single-layer model (SLM; circles) and the plasmasphere extended SLM (IRI-Plas; triangles). For reference, we include the RM reported in the ATNF pulsar catalogue from \citet{Johnston2005} (black dashed line) and the corresponding uncertainty estimate (grey shaded region).}
    \label{fig:J0630_short}
\end{figure}

In Figure \ref{fig:J0630_short}, we present the observed and ionosphere-subtracted RMs for nine observations of \control{} over a $\sim$6\,h time period on UT+8 2022-10-08.
The first four observations were collected before local sunrise, and show small variations of the order $\sim$0.01\,\rmu{} (see the inset in Figure \ref{fig:J0630_short}).
After sunrise, we observe the expected downward trend in \RMobs{} due to the diurnal photoionisation of the ionosphere.
The diurnal trend is the largest source of potential error in the estimated RM of the ISM.
Therefore, it is essential that this modulation is properly subtracted.
As a simple test, we compare the residual RMs after subtracting the estimated \RMion{} from two GIMs (JPLG and UQRG) with and without scaling using the IRI-Plas ionospheric density profile (see Figure \ref{fig:J0630_short}, bottom panel).
For reference, we also show the most up to date RM in the ATNF pulsar catalogue \citep[from][]{Johnston2005}.

Since the ISM does not vary appreciably on the timescale of hours, we expect any temporal variations in $\mathrm{RM}_\mathrm{obs}-\mathrm{RM}_\mathrm{ion}$ (the `residual' RM) on this timescale to be due to the imperfect subtraction of ionospheric variability.
Additionally, because \control{} lies outside of the Gum Nebula, it is not expected show any long-term change in RM.
Variations in the residual RM on longer timescales could therefore also be due to the ionospheric model.
To quantify these variations, we use the mean absolute deviation\footnote{We use the \textit{mean} absolute deviation rather than the \textit{median} absolute deviation so that models with outliers are unfavoured.} from the mean of the data.
We find that the RM residuals from the single-layer model deviate less than the estimates made using the plasmasphere-extended model (i.e., the single-layer model scaled by the IRI-Plas density profile).
Furthermore, we see greater deviations during the daytime hours, especially for the plasmasphere-extended model (see Figure \ref{fig:J0630_short}).
The trend suggests that the plasmasphere-extended model overestimates the ionospheric contribution.
Between the two GIMs, we find that JPLG deviates less overall than UQRG, which is due to several significant outliers in the UQRG estimates.
The outliers from the UQRG GIM may be a result of the poor GNSS receiver coverage at the MRO, which we discuss further in \S\ref{sec:implications}.
Interestingly, UQRG shows a lower deviation than JPLG when the outliers are excluded.
Nevertheless, we use JPLG for all further analysis as it appears to yield more consistent results.

\subsection{Temporal RM variations}\label{sec:rm}

\begin{figure*}
    \centering
    \includegraphics[width=0.484\linewidth]{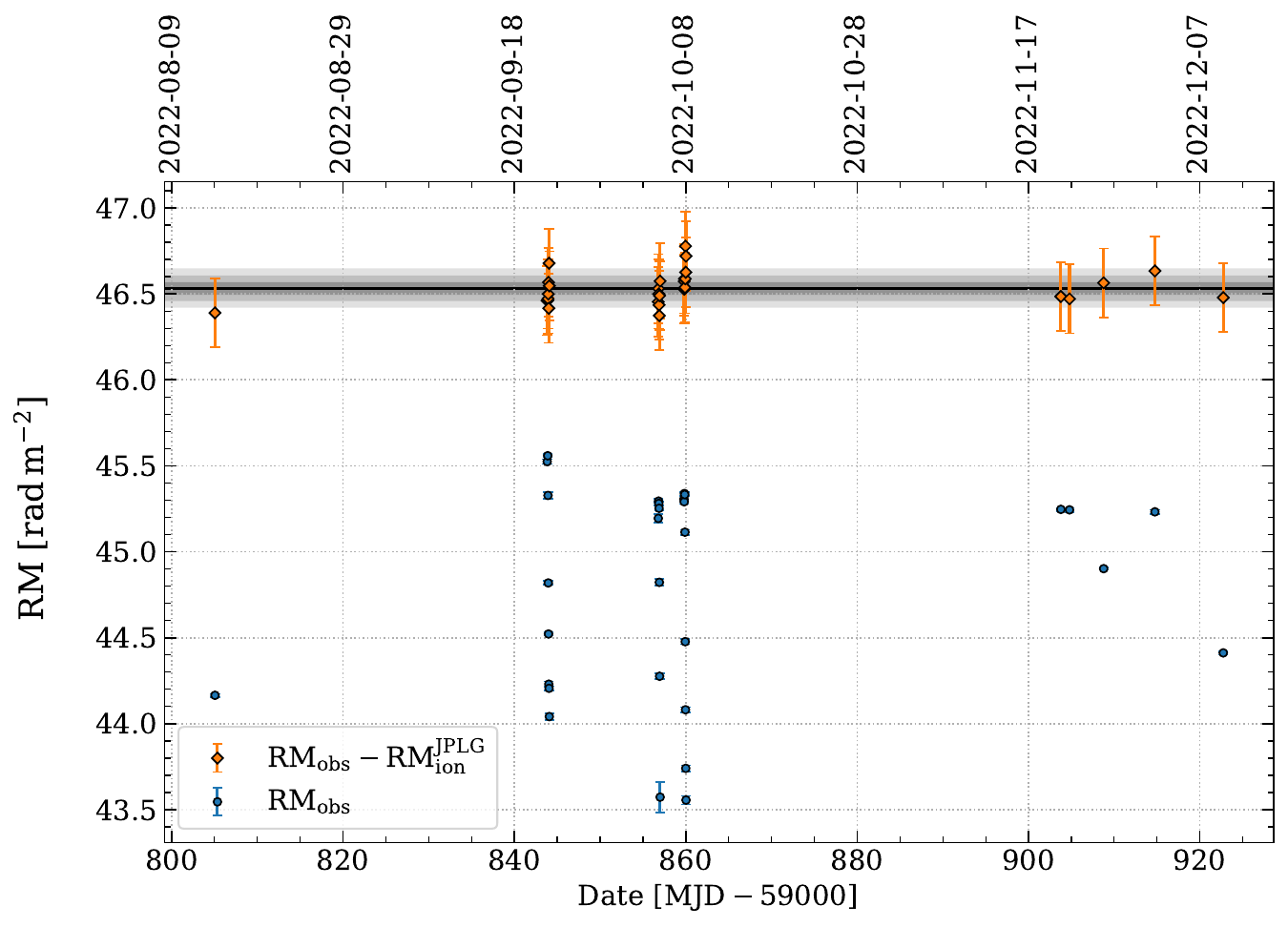}
    \quad
    \includegraphics[width=0.49\linewidth]{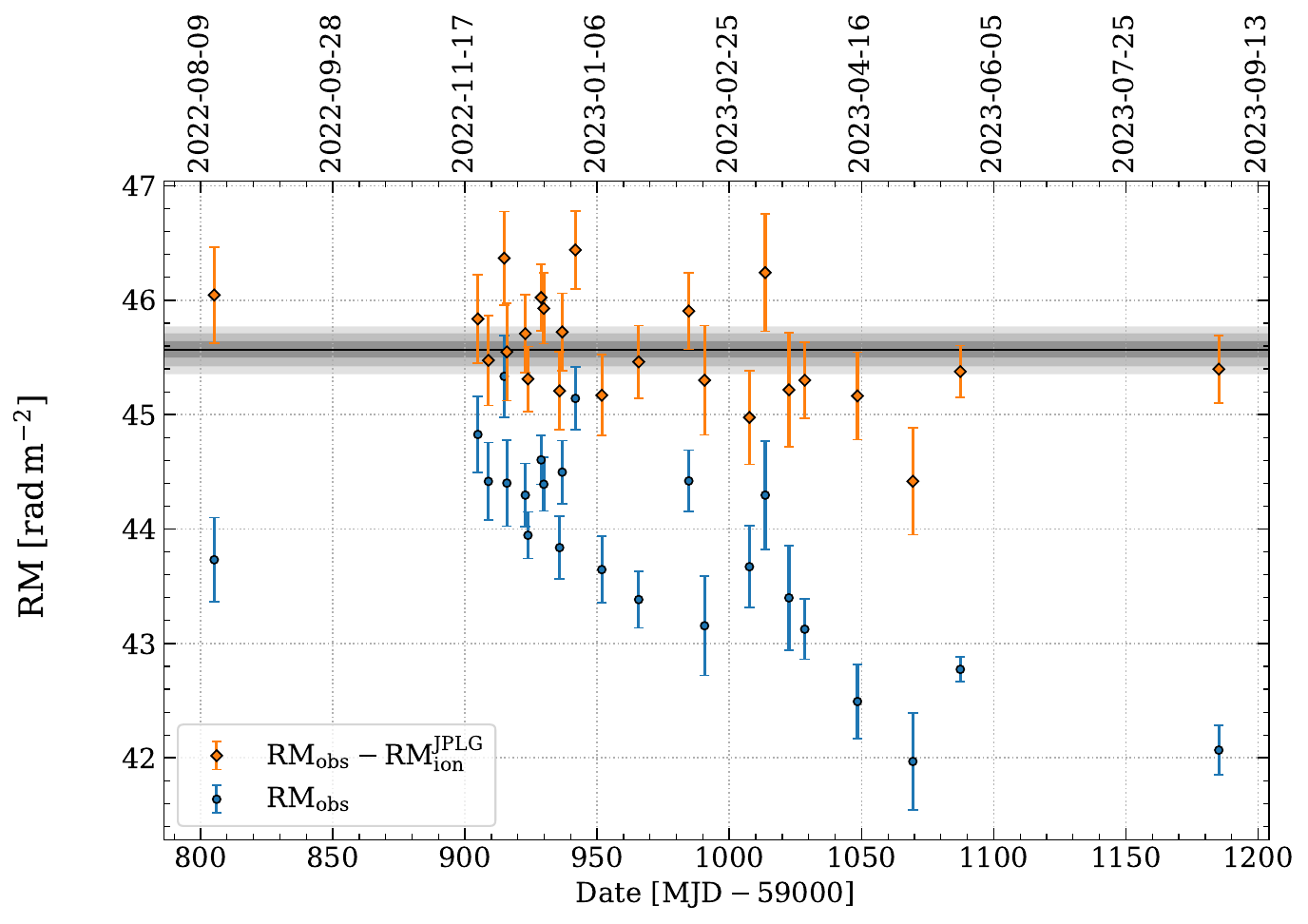}
    \caption{RM time series for \control{} (left) and \vela{} (right). Each plot shows the RM before (blue circles) and after (orange diamonds) subtracting the ionospheric RM. We show the weighted mean (black line) of the ionosphere-corrected RM, and the 1, 2, and 3$\sigma$ credible intervals of the posterior probability distribution (grey shaded bands).}
    \label{fig:rm_long}
\end{figure*}

In Figure \ref{fig:rm_long}, we show the complete RM time series for each pulsar before and after subtracting the JPLG/SLM model.
To determine the best estimate of \RMism, we measure the weighted mean of $\mathrm{RM}_\mathrm{obs}-\mathrm{RM}_\mathrm{ion}$ by fitting a constant (time-independent) model with a Gaussian likelihood.
We set a uniform prior probability and used the Bayesian inference library \texttt{Bilby} to estimate the posterior probability distribution \citep{Ashton2019}.
The median and 1, 2, and 3$\sigma$ credible intervals of the model fit are shown in Figure \ref{fig:rm_long}.

For \control, we measure a weighted mean of $46.53\pm 0.04$\,\rmu{} (where the uncertainty is the 1$\sigma$ credible interval), which is in excellent agreement with the value of $46.53\pm0.12$\,\rmu{} from \citet{Johnston2005}.
The RM uncertainties all overlap with the weighted mean of the data, which is indicative that the uncertainties may be overestimated.
However, the conservatively large uncertainties account for the lack of knowledge about the accuracy of ionospheric models at the MRO.
A linear model fit to the data shows that the gradient is consistent with zero to within 1$\sigma$.

For \vela, the observed RM decreases over time due to the shift in transit time (and thus the observing time) throughout the year.
After subtracting the JPLG/SLM model, we measure a weighted mean RM of $45.57\pm 0.07$\,\rmu, which is consistent with the value of $45.3\pm 0.7$\,\rmu{} from \citet{Sobey2021}.
We also observe a residual gradient of $-0.8\pm0.3$\,\rmuyr, which could be the result of a systematic error in the ionospheric model.
For example, \citet{Porayko2019} observed a systematic error with a period of 1\,d in RM residuals after subtracting the JPLG model.
This could cause a corresponding error with a period of 1\,yr in our data, due to the observations being collected at progressively earlier times of day.
An error of this kind would be less significant in the RM series of \control, which spans only $\sim$0.3\,yr compared to $\sim$1\,yr for Vela, and has sparser temporal coverage.
If instead we subtract the UQRG/SLM model, we find that the gradient is consistent with zero.
The fact that the long-term trend is model dependent is evidence that the trend is systematic in origin, caused by inaccuracies in the JPLG model on an annual timescale.
Observations over multiple years will be required to disentangle model systematics from the underlying astrophysical trend.

\subsection{Temporal DM variations}
\begin{figure}[t]
    \centering
    \includegraphics[width=\linewidth]{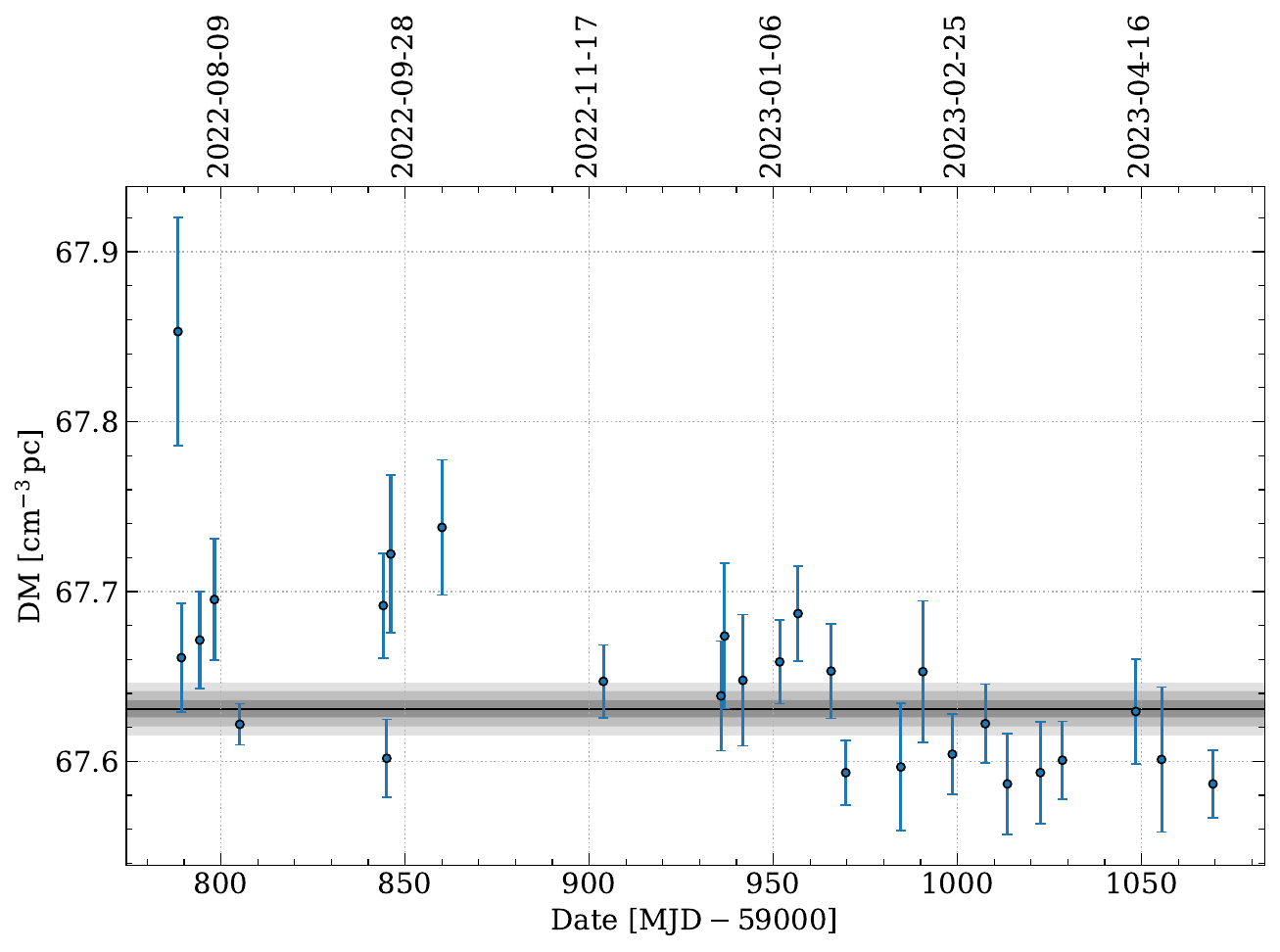}
    \caption{DM time series for \vela. We show the weighted mean (black line) and the 1, 2, and 3$\sigma$ credible intervals of the posterior probability distribution (grey shaded bands).}
    \label{fig:dm}
\end{figure}

In Figure \ref{fig:dm}, we show the measured DMs for \vela{} from observations at 326.17\,MHz.
As in \S\ref{sec:rm}, we fit a constant model with a Gaussian likelihood, and measure a weighted mean value of $67.63\pm 0.01$\,\dmu.
The time series shows outliers and correlated variability that could be attributed to temporal variations in the pulse broadening time.
Such variations can bias the pulse arrival times due to discrepancies between the template profile and the individual observations.
A detailed analysis of these variations, including the influence of scattering on the DM measurements, is beyond the scope of this work, and will be deferred to a future publication.

\section{Discussion}\label{sec:discussion}
\subsection{Long-term RM and DM variations}
The RM and DM of the Vela pulsar have long been observed to change significantly over time, initially showing a correlated linear trend which was attributed to the relative movement of the pulsar and a compact overdensity (filament) of magnetised plasma within the Vela SNR \citep{Hamilton1977}.
Later analysis of RM measurements published over several decades revealed a more complex deterministic trend; \citet{XueThesis} suggest that the variations in RM could be explained by an inhomogeneous magnetic field within the plasma filament.

\begin{figure}[t]
    \centering
    \includegraphics[width=\linewidth]{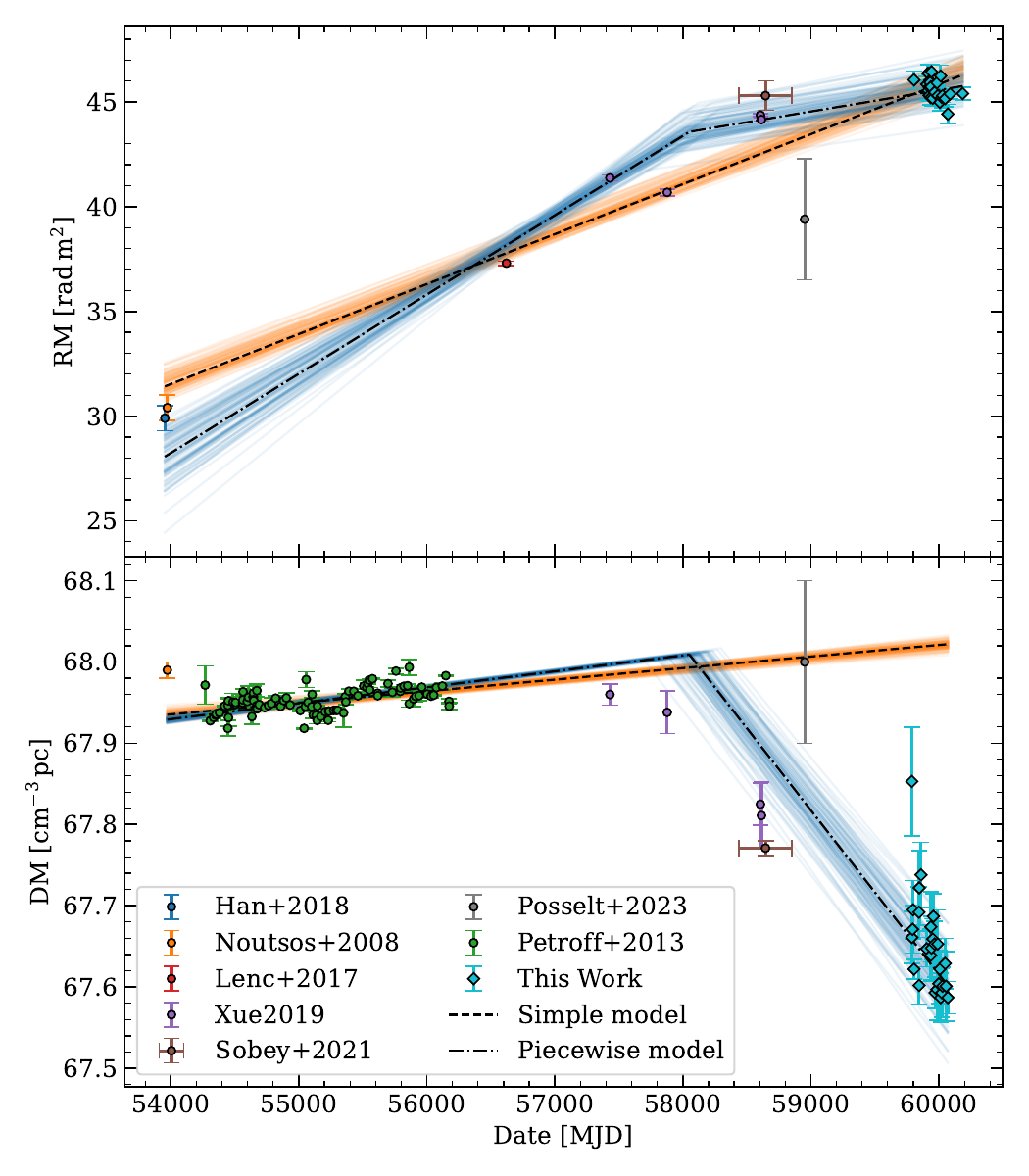}
    \caption{Published measurements of RM (top) and DM (bottom) for \vela{} between 2006 and 2023. The black dashed lines show the median simple linear models and the black dot-dashed lines show the median piecewise linear models. The thin orange and blue lines show 100 random samples from the posterior distributions of the simple and piecewise model fits, respectively. The diamond markers show measurements obtained using the AAVS2 in this work; the circle markers show data from \citet{Noutsos2008,Petroff2013,Lenc2017,Han2018,XueThesis,Sobey2021,Posselt2023}. We show horizontal error bars for \citet{Sobey2021} as the measurements are averaged over multiple observations (the horizontal uncertainties were not included in the model fits).}
    \label{fig:lit}
\end{figure}

In this section, we focus on the most recent two decades of RM and DM variations for the Vela pulsar, which is yet to be analysed in detail in the literature.
In Figure \ref{fig:lit}, we plot all published RM and DM measurements since 2006, including the measurements from this work.
From visual inspection of the data, it is immediately apparent that the DM decreased by $\sim$0.3--0.4\,\dmu{} between MJD $\sim$56000 \citep{Petroff2013} and MJD $\sim$60000 (this work).
The sparsity of published data between these dates makes it difficult to disentangle the true change in astrophysical DM with systematic differences due to the profile evolution with frequency and measurement methods.
However, the agreement in DM between Murriyang observations at $\sim$700--4000\,MHz \citep{Sobey2021} and MWA observations at $\sim$170--230\,MHz \citep{XueThesis} collected on similar observing epochs (MJD $\sim$58400--58900) suggests that the systematic offset arising from the difference in observing frequencies may not be significant to first order.
With this in mind, the recent decrease in DM suggests that, contrary to the conclusions of \citet{Petroff2013}, the filament seen by \citet{Hamilton1977} may still be moving out of the line of sight.

Following previous analysis in the literature \citep[e.g.][]{Hamilton1977,Petroff2013,XueThesis}, we fit linear models to measure the gradients in DM and RM over time.
Specifically, we consider a simple (one-segment) and a piecewise (two-segment) linear model.
Since the RM and DM both depend on the thermal electron density along the line of sight, a significant change in density could result in a correlated change in both quantities.
To account for this, we have enforced that the break point in the piecewise model is the same for both the RM and DM.
We jointly fit the RM and DM data using a Gaussian likelihood, and make use of the Huber loss function to reduce the weighting of outliers greater than 1.345$\sigma$ from the model (see \ref{app:model} for further details about the model fitting procedure).
This approach results in a model fit which is less constrained by measurements with underestimated uncertainties or large systematic offsets from the rest of the data.

In Table \ref{tab:litfit}, we present estimates of the RM and DM gradients and the magnetic field strength computed using equation \eqref{eq:Bfil}.
We estimate each parameter using the median of the posterior distribution, with the uncertainties indicating the 68\% credible interval.
From the simple linear model fit, we measure $\langle B_\parallel\rangle_\mathrm{var} = 206^{+23}_{-19}\,\mu\mathrm{G}$.
The piecewise model fit deviates significantly from the simple model, suggesting that the mean magnetic field in the time-varying region transitions from $240^{+30}_{-20}\,\mu\mathrm{G}$ to $-6.2^{+0.7}_{-0.9}\,\mu\mathrm{G}$ at around MJD $58050^{+90}_{-100}$.
The Bayes factor (i.e, the difference in evidence) between the two model fits is $\log\mathcal{B}=535.9\pm 0.2$, which strongly favours the piecewise model.
The sign change implies that the mean direction of the magnetic field changes from pointing towards the observer to pointing towards the pulsar.
Whilst the estimated magnetic field strengths exceed previous estimates for the Vela pulsar \citep[$\sim$20--90\,$\mu\mathrm{G}$;][]{Hamilton1985,XueThesis}, they are not unreasonable for SNRs.
For example, \citet{Rankin1988} measured $\langle B_\parallel\rangle_\mathrm{var}=170\,\mu\mathrm{G}$ for the Crab pulsar.

Following \citet{Hamilton1977}, we can estimate the transverse scale length of the pulsar's path through the ISM as $\ell = v_\perp \Delta t$, where $v_\perp$ is the transverse space velocity in $\mathrm{pc}\,\mathrm{yr}^{-1}$ and $\Delta t$ is the time span in years.
Assuming a similar scale in the line-of-sight dimension, the excess electron density in the time-varying region (i.e., the change in electron density) is $\Delta n_e = (\mathrm{dDM}/\mathrm{d}t) v_\perp^{-1}$.
The Vela pulsar has a measured transverse velocity of $61\pm 2\,\mathrm{km}\,\mathrm{s}^{-1}$ \citep{Dodson2003}, or equivalently, $(6.2\pm0.2) \times 10^{-7}\,\mathrm{pc}\,\mathrm{yr}^{-1}$, placing it at the low end of the velocity distribution for young pulsars \citep[e.g.,][]{Lyne1994}.
The derived measurements are included in Table \ref{tab:litfit}.
We find that the scale lengths of the time-varying regions in the model fits are around $\sim$1--2\,\textsc{au}.
From the simple model fit, we measure an increase in density of $(8.3\pm 0.8)\times 10^3\,\mathrm{cm}^{-3}$.
The piecewise model shows a similar density increase before the break point, followed by a steep decrease of $(1.18\pm 0.08)\times 10^5\,\mathrm{cm}^{-3}$ after the break point.
These density measurements are at least 1--2 orders of magnitude greater than some estimates for bright filaments in the Crab nebula \citep{Osterbrock1957,Woltjer1958}.
Interestingly, the observed structure is comparable in density and length scale with discrete compact structures associated with extreme scattering events in the ISM \citep[e.g.,][]{Fiedler1987,Clegg1998,Dong2018}, which may reside in old supernova remnants \citep{Romani1987}.
Of course, another way to reconcile the high implied densities is to consider a structure which is significantly elongated along the line of sight, in which case our measurements would be overestimated.
In any case, further observations will be required to better understand the origin of the structure.

One potential cause for a large magnetic field gradient is a compression due to shocks in the ISM.
Assuming a simple spherical compression, the magnetic field strength and electron density are positively correlated such that $B\propto n_e^{2/3}$ \citep[e.g.,][]{Seta2022}.
In this case, the decrease in magnetic field strength derived from the piecewise model fit corresponds to a density decrease by a factor of $\sim$240.
For a decrease in density of $10^5\,\mathrm{cm}^{-3}$, the final density in the time-varying region is $\sim$$420\,\mathrm{cm}^{-3}$.
It is important to note that the power-law relation is an idealised model and the true relation is more complex in a turbulent medium \citep{Seta2022}.

In the above analysis, we have assumed that the time-varying region is a single compact structure.
It is also possible that the observed variations are the result of multiple structures at different locations along the line of sight.
In principle, this could be tested by observing interstellar scintillation in Vela pulsar observations at higher frequencies, which can be used to determine the location of one or multiple scattering screens \citep[e.g.,][]{Reardon2020}.
Alternatively, this could be tested via accurate modelling of the pulse shapes at low observing frequencies ($\lesssim$300\,MHz), which may necessitate reliably disentangling the profile evolution (in frequency) and frequency-dependent effects of pulse broadening that arise from multipath scattering \citep[e.g.,][]{Bhat2004,Geyer2017}.

\begin{table*}
    \centering
    \caption{Estimated RM and DM gradients and the implied mean magnetic field strength of the magnetised plasma filament. The first row shows the results from a simple linear model; the second and third rows show the results from a piecewise linear model with a variable break point. The columns (from left to right) are: the start and end dates of the linear segment; the transverse scale size probed by the linear model ($\ell$); the RM gradient ($\mathrm{dRM}/\mathrm{d}t$); the DM gradient ($\mathrm{dDM}/\mathrm{d}t$); the mean magnetic field of the time-varying region parallel to the line of sight (\Bparvar); and the change in electron density ($\Delta n_e$). See text for details.}
    \label{tab:litfit}
    \begin{tabular}{ccccccc}
        \hline
        MJD start     & MJD end       & $\ell$         &$\mathrm{dRM}/\mathrm{d}t$ & $\mathrm{dDM}/\mathrm{d}t$ & \Bparvar             & $\Delta n_e$ \\
        \text{[}day]  & [day]         & [\textsc{au}]  & [\rmuyr]                  & [$10^{-3}$\,\dmuyr]        & [$\mu\mathrm{G}$]    & [$10^3\,\mathrm{cm}^{-3}$] \\
        \hline
        $53953$       & $60185$       & $2.20\pm 0.07$ & $0.87^{+0.03}_{-0.04}$    & $5.2\pm 0.5$        & $206^{+23}_{-19}$    & $8.3\pm 0.8$ \\
        \hline
        $53953$       & $58050^{+90}_{-100}$ & $1.44\pm 0.06$         & $1.38^{+0.14}_{-0.11}$    & $7.2\pm 0.2$     & $240^{+30}_{-20}$    & $11.5\pm 0.5$ \\
        $58050^{+90}_{-100}$ & $60185$       & $0.75^{+0.05}_{-0.04}$ & $0.37^{+0.04}_{-0.03}$    & $-74\pm 4$       & $-6.2^{+0.7}_{-0.9}$ & $-118\pm 8$ \\
        \hline
    \end{tabular}
\end{table*}

\subsection{Implications for future low-frequency pulsar monitoring}\label{sec:implications}
The results of this work, and future studies of the magneto-ionised ISM using the SKA-Low, could be improved with more accurate ionospheric models.
Current models are limited by the fact that there is currently only one GNSS ground station at the MRO, with the next nearest stations being 290 and 486\,km away.
For comparison, the EUREF Permanent GNSS Network\footnote{\url{https://epncb.oma.be/_networkdata/stationmaps.php}} comprises over 400 GNSS stations distributed throughout Western Europe; $\sim$10 of these are within 200\,km of the LOFAR core.
Having a network of GNSS stations throughout and surrounding the MRO would result in data from more pierce points, which would improve the spacial and temporal resolution of TEC models.
In principle, it is also possible to use a widefield interferometer (such as the MWA) to measure refractive shifts in compact radio sources \citep[e.g.,][]{Jordan2017}, and use these to improve TEC estimates from global models.
For monitoring campaigns with multiple target pulsars, calibration observations could be scheduled between pulsar observations to improve the temporal resolution of the TEC estimates.
The Phase III capabilities of the MWA may enable this observing strategy.
Additionally, the efficiency of pulsar monitoring will be significantly improved by the subarray capabilities of the SKA-Low, which could enable simultaneous multi-beamforming and imaging to obtain precise RMs towards a collection of pulsars.
The Gum Nebula is a particularly strong science case; as demonstrated by \citet{XueThesis}, polarimetric pulsar observations can be used to probe the size and structure of the Nebula.

\section{Conclusions}\label{sec:conclusions}
We have analysed low-frequency observations of the Vela pulsar and PSR \control{} collected with the upgraded bandwidth of the AAVS2 prototype station.
We find that the polarimetric pulse profiles are consistent with published detections from the MWA, the GMRT, and Murriyang, which is the first validation of pulsar polarimetry from an SKA-Low precursor station.
We used the pulsar detections to obtain precise RM estimates for both pulsars, as well as DM estimates for Vela.
From a basic comparison of several ionospheric models, we find that the best ionospheric RM estimates come from the JPLG model with a single-layer approximation for the ionospheric shell.

From our observations of Vela, spanning over 1\,yr, we observe a notable gradient in the RM; however, the temporal sampling is too sparse to confirm an astrophysical origin.
More likely, this gradient originates from a systematic error in the ionospheric model.
We do not find any significant gradient in RM for PSR \control{}, although the shorter time span of the data ($\sim$0.3\,yr) reduces the sensitivity to annual systematics (i.e., those with a period of 1\,yr).
Further improvements to ionospheric models and a longer-term, higher-cadence observing campaign will be necessary to disentangle systematics from ISM variations on these timescales ($\sim$months).

By combining our RM and DM data for Vela with published data from 2006 to present, we detect a significant change in DM, and a less significant but correlated change in RM.
The temporal variations in this data set probe transverse scale lengths of $\sim$1--2\,\textsc{au}.
We detect a change in the magnetic field strength within the time-varying region from $240^{+30}_{-20}\,\mu\mathrm{G}$ to $-6.2^{+0.7}_{-0.9}\,\mu\mathrm{G}$, and a corresponding reduction in density by $\sim$$10^5\,\mathrm{cm}^{-3}$.
These measurements suggest the presence of an extremely dense and compact plasma structure, comparable to those associated with extreme scattering events.
Further monitoring of RM and DM will provide a useful data set to further investigate microstructure in the ISM towards the Vela pulsar.

Future studies of the ISM using pulsar monitoring could be a niche science opportunity for the MWA and SKA-Low, yielding interesting results from relatively little telescope time.
This work highlights the potential for SKA-Low stations to be used for monitoring science, even in the early stages of constructing the SKA-Low.
The precision of these studies is currently limited by poorly constrained models of the highly dynamic ionosphere.
Further work will be necessary to improve the accuracy of ionospheric RM calibration, either through radio imaging or by expanding the network of GNSS receivers at the MRO.

\begin{acknowledgement}
We thank Dejan Uro{\v{s}}evi{\'c} for a constructive review which helped to improve this manuscript.
We also thank Amit Seta and Poonam Chandra for helpful discussions, Maaijke Mevius for providing ionospheric RM estimates, and Emily Petroff for providing the DM data from \citet{Petroff2013}.
We acknowledge Riccardo Chiello for his contributions to the TPM firmware.
C.P.L. was supported by an Australian Government Research Training Program (RTP) Stipend and RTP Fee-Offset Scholarship.
This scientific work makes use of Inyarrimanha Ilgari Bundara, CSIRO's Murchison Radio-astronomy Observatory.
We acknowledge the Wajarri Yamatji people as the traditional owners of the Observatory site. 
This work was supported by resources provided by the Pawsey Supercomputing Research Centre with funding from the Australian Government and the Government of Western Australia. 
The AAVS2 is hosted by the MWA under an agreement via the MWA External Instruments Policy. The acquisition system was designed and purchased by INAF/Oxford University and the RX chain was design by INAF, as part of the SKA design and prototyping program.
We acknowledge the support of the Curtin operations team, INAF group, and SKA-Low team in the development and on-going maintenance of the facilities used in this work.
This work made use of NASA's Astrophysics Data System and arXiv.

Software/packages:
\texttt{Astropy} \citep{AstroPy},
\texttt{Bilby} \citep{Ashton2019},
\texttt{corner} \citep{corner},
\texttt{DSPSR} \citep{vanStraten2011},
\texttt{Dynesty} \citep{Speagle2020},
\texttt{Matplotlib} \citep{Matplotlib},
\texttt{Numpy} \citep{NumPy},
\texttt{PSRCHIVE} \citep{Hotan2004,vanStraten2012},
\texttt{PSRSALSA} \citep{Weltevrede2016},
\texttt{RMextract} \citep{RMextract},
\texttt{RMtoolkit} \citep{Heald2009},
\texttt{Tempo2} \citep{Hobbs2006}.
\end{acknowledgement}

\printbibliography

@INPROCEEDINGS{vanEs2020,
       author = {{van Es}, A.~J.~J. and {Labate}, M.~G. and {Waterson}, M.~F. and {Monari}, J. and {Bolli}, P. and {Davidson}, D. and {Wayth}, R. and {Sokolowski}, M. and {Di Ninni}, P. and {Pupillo}, G. and {Macario}, G. and {Virone}, G. and {Ciorba}, L. and {Paonessa}, F.},
        title = "{A prototype model for evaluating SKA-LOW station calibration}",
    booktitle = "{Ground-based and Airborne Telescopes VIII}",
         year = 2020,
       editor = {{Marshall}, Heather K. and {Spyromilio}, Jason and {Usuda}, Tomonori},
       series = {Society of Photo-Optical Instrumentation Engineers (SPIE) Conference Series},
       volume = {11445},
        month = dec,
          eid = {1144589},
        pages = {1144589},
          doi = {10.1117/12.2562391},
       adsurl = {https://ui.adsabs.harvard.edu/abs/2020SPIE11445E..89V}
}

@ARTICLE{Macario2022,
       author = {{Macario}, Giulia and {Pupillo}, Giuseppe and {Bernardi}, Gianni and {Bolli}, Pietro and {Di Ninni}, Paola and {Comoretto}, Giovanni and {Mattana}, Andrea and {Monari}, Jader and {Perini}, Federico and {Schiaffino}, Marco and {Sokolowski}, Marcin and {Wayth}, Randall and {Broderick}, Jess and {Waterson}, Mark and {Grazia Labate}, Maria and {Chiello}, Riccardo and {Magro}, Alessio and {Booler}, Tom and {Mcphail}, Andrew and {Minchin}, Dave and {Bhushan}, Raunaq},
        title = "{Characterization of the SKA1-Low prototype station Aperture Array Verification System 2}",
      journal = {Journal of Astronomical Telescopes, Instruments, and Systems},
         year = 2022,
        month = jan,
       volume = {8},
          eid = {011014},
        pages = {011014},
          doi = {10.1117/1.JATIS.8.1.011014},
archivePrefix = {arXiv},
       eprint = {2109.11983},
 primaryClass = {astro-ph.IM},
       adsurl = {https://ui.adsabs.harvard.edu/abs/2022JATIS...8a1014M}
}

@ARTICLE{Lee2022,
       author = {{Lee}, C.~P. and {Bhat}, N.~D.~R. and {Sokolowski}, M. and {Swainston}, N.~A. and {Ung}, D. and {Magro}, A. and {Chiello}, R.},
        title = "{Spectral analysis of 22 radio pulsars using SKA-Low precursor stations}",
      journal = {\pasa},
         year = 2022,
        month = sep,
       volume = {39},
          eid = {e042},
        pages = {e042},
          doi = {10.1017/pasa.2022.40},
archivePrefix = {arXiv},
       eprint = {2208.07182},
 primaryClass = {astro-ph.HE},
       adsurl = {https://ui.adsabs.harvard.edu/abs/2022PASA...39...42L}
}

@ARTICLE{vanStraten2011,
       author = {{van Straten}, W. and {Bailes}, M.},
        title = "{DSPSR: Digital Signal Processing Software for Pulsar Astronomy}",
      journal = {\pasa},
         year = 2011,
        month = jan,
       volume = {28},
       number = {1},
        pages = {1-14},
          doi = {10.1071/AS10021},
archivePrefix = {arXiv},
       eprint = {1008.3973},
 primaryClass = {astro-ph.IM},
       adsurl = {https://ui.adsabs.harvard.edu/abs/2011PASA...28....1V}
}

@ARTICLE{Hotan2004,
       author = {{Hotan}, A.~W. and {van Straten}, W. and {Manchester}, R.~N.},
        title = "{PSRCHIVE and PSRFITS: An Open Approach to Radio Pulsar Data Storage and Analysis}",
      journal = {\pasa},
         year = 2004,
        month = jan,
       volume = {21},
       number = {3},
        pages = {302-309},
          doi = {10.1071/AS04022},
archivePrefix = {arXiv},
       eprint = {astro-ph/0404549},
 primaryClass = {astro-ph},
       adsurl = {https://ui.adsabs.harvard.edu/abs/2004PASA...21..302H}
}

@ARTICLE{vanStraten2012,
       author = {{van Straten}, Willem and {Demorest}, Paul and {Oslowski}, Stefan},
        title = "{Pulsar Data Analysis with PSRCHIVE}",
      journal = {Astronomical Research and Technology},
         year = 2012,
        month = jul,
       volume = {9},
       number = {3},
        pages = {237-256},
          doi = {10.48550/arXiv.1205.6276},
archivePrefix = {arXiv},
       eprint = {1205.6276},
 primaryClass = {astro-ph.IM},
       adsurl = {https://ui.adsabs.harvard.edu/abs/2012AR&T....9..237V}
}

@ARTICLE{Han2018,
       author = {{Han}, J.~L. and {Manchester}, R.~N. and {van Straten}, W. and {Demorest}, P.},
        title = "{Pulsar Rotation Measures and Large-scale Magnetic Field Reversals in the Galactic Disk}",
      journal = {\apjs},
         year = 2018,
        month = jan,
       volume = {234},
       number = {1},
          eid = {11},
        pages = {11},
          doi = {10.3847/1538-4365/aa9c45},
archivePrefix = {arXiv},
       eprint = {1712.01997},
 primaryClass = {astro-ph.GA},
       adsurl = {https://ui.adsabs.harvard.edu/abs/2018ApJS..234...11H}
}

@phdthesis{XueThesis,
   author = {Mengyao Xue},
   institution = {Curtin Institute of Radio Astronomy},
   school = {Curtin University},
   month = nov,
   title = "{Characterising the Local Interstellar Medium using Low-Frequency Pulsar Polarimetry}",
   year = {2019},
   url = {http://hdl.handle.net/20.500.11937/80645}
}

@ARTICLE{Jones2017,
       author = {{Jones}, M.~L. and {McLaughlin}, M.~A. and {Lam}, M.~T. and {Cordes}, J.~M. and {Levin}, L. and {Chatterjee}, S. and {Arzoumanian}, Z. and {Crowter}, K. and {Demorest}, P.~B. and {Dolch}, T. and {Ellis}, J.~A. and {Ferdman}, R.~D. and {Fonseca}, E. and {Gonzalez}, M.~E. and {Jones}, G. and {Lazio}, T.~J.~W. and {Nice}, D.~J. and {Pennucci}, T.~T. and {Ransom}, S.~M. and {Stinebring}, D.~R. and {Stairs}, I.~H. and {Stovall}, K. and {Swiggum}, J.~K. and {Zhu}, W.~W.},
        title = "{The NANOGrav Nine-year Data Set: Measurement and Analysis of Variations in Dispersion Measures}",
      journal = {\apj},
         year = 2017,
        month = jun,
       volume = {841},
       number = {2},
          eid = {125},
        pages = {125},
          doi = {10.3847/1538-4357/aa73df},
archivePrefix = {arXiv},
       eprint = {1612.03187},
 primaryClass = {astro-ph.HE},
       adsurl = {https://ui.adsabs.harvard.edu/abs/2017ApJ...841..125J}
}

@ARTICLE{Wahl2022,
       author = {{Wahl}, H.~M. and {McLaughlin}, M.~A. and {Gentile}, P.~A. and {Jones}, M.~L. and {Spiewak}, R. and {Arzoumanian}, Z. and {Crowter}, K. and {Demorest}, P.~B. and {DeCesar}, M.~E. and {Dolch}, T. and {Ellis}, J.~A. and {Ferdman}, R.~D. and {Ferrara}, E.~C. and {Fonseca}, E. and {Garver-Daniels}, N. and {Jones}, G. and {Lam}, M.~T. and {Levin}, L. and {Lewandowska}, N. and {Lorimer}, D.~R. and {Lynch}, R.~S. and {Madison}, D.~R. and {Ng}, C. and {Nice}, D.~J. and {Pennucci}, T.~T. and {Ransom}, S.~M. and {Ray}, P. and {Stairs}, I.~H. and {Stovall}, K. and {Swiggum}, J.~K. and {Zhu}, W.~W.},
        title = "{The NANOGrav 12.5 yr Data Set: Polarimetry and Faraday Rotation Measures from Observations of Millisecond Pulsars with the Green Bank Telescope}",
      journal = {\apj},
         year = 2022,
        month = feb,
       volume = {926},
       number = {2},
          eid = {168},
        pages = {168},
          doi = {10.3847/1538-4357/ac4045},
archivePrefix = {arXiv},
       eprint = {2104.05723},
 primaryClass = {astro-ph.SR},
       adsurl = {https://ui.adsabs.harvard.edu/abs/2022ApJ...926..168W}
}

@ARTICLE{You2007,
       author = {{You}, X.~P. and {Hobbs}, G. and {Coles}, W.~A. and {Manchester}, R.~N. and {Edwards}, R. and {Bailes}, M. and {Sarkissian}, J. and {Verbiest}, J.~P.~W. and {van Straten}, W. and {Hotan}, A. and {Ord}, S. and {Jenet}, F. and {Bhat}, N.~D.~R. and {Teoh}, A.},
        title = "{Dispersion measure variations and their effect on precision pulsar timing}",
      journal = {\mnras},
         year = 2007,
        month = jun,
       volume = {378},
       number = {2},
        pages = {493-506},
          doi = {10.1111/j.1365-2966.2007.11617.x},
archivePrefix = {arXiv},
       eprint = {astro-ph/0702366},
 primaryClass = {astro-ph},
       adsurl = {https://ui.adsabs.harvard.edu/abs/2007MNRAS.378..493Y}
}

@ARTICLE{Sotomayor-Beltran2013,
       author = {{Sotomayor-Beltran}, C. and {Sobey}, C. and {Hessels}, J.~W.~T. and {de Bruyn}, G. and {Noutsos}, A. and {Alexov}, A. and {Anderson}, J. and {Asgekar}, A. and {Avruch}, I.~M. and {Beck}, R. and {Bell}, M.~E. and {Bell}, M.~R. and {Bentum}, M.~J. and {Bernardi}, G. and {Best}, P. and {Birzan}, L. and {Bonafede}, A. and {Breitling}, F. and {Broderick}, J. and {Brouw}, W.~N. and {Br{\"u}ggen}, M. and {Ciardi}, B. and {de Gasperin}, F. and {Dettmar}, R. -J. and {van Duin}, A. and {Duscha}, S. and {Eisl{\"o}ffel}, J. and {Falcke}, H. and {Fallows}, R.~A. and {Fender}, R. and {Ferrari}, C. and {Frieswijk}, W. and {Garrett}, M.~A. and {Grie{\ss}meier}, J. and {Grit}, T. and {Gunst}, A.~W. and {Hassall}, T.~E. and {Heald}, G. and {Hoeft}, M. and {Horneffer}, A. and {Iacobelli}, M. and {Juette}, E. and {Karastergiou}, A. and {Keane}, E. and {Kohler}, J. and {Kramer}, M. and {Kondratiev}, V.~I. and {Koopmans}, L.~V.~E. and {Kuniyoshi}, M. and {Kuper}, G. and {van Leeuwen}, J. and {Maat}, P. and {Macario}, G. and {Markoff}, S. and {McKean}, J.~P. and {Mulcahy}, D.~D. and {Munk}, H. and {Orru}, E. and {Paas}, H. and {Pandey-Pommier}, M. and {Pilia}, M. and {Pizzo}, R. and {Polatidis}, A.~G. and {Reich}, W. and {R{\"o}ttgering}, H. and {Serylak}, M. and {Sluman}, J. and {Stappers}, B.~W. and {Tagger}, M. and {Tang}, Y. and {Tasse}, C. and {ter Veen}, S. and {Vermeulen}, R. and {van Weeren}, R.~J. and {Wijers}, R.~A.~M.~J. and {Wijnholds}, S.~J. and {Wise}, M.~W. and {Wucknitz}, O. and {Yatawatta}, S. and {Zarka}, P.},
        title = "{Calibrating high-precision Faraday rotation measurements for LOFAR and the next generation of low-frequency radio telescopes}",
      journal = {\aap},
         year = 2013,
        month = apr,
       volume = {552},
          eid = {A58},
        pages = {A58},
          doi = {10.1051/0004-6361/201220728},
archivePrefix = {arXiv},
       eprint = {1303.6230},
 primaryClass = {astro-ph.IM},
       adsurl = {https://ui.adsabs.harvard.edu/abs/2013A&A...552A..58S}
}

@ARTICLE{Hamilton1977,
       author = {{Hamilton}, P.~A. and {McCulloch}, P.~M. and {Manchester}, R.~N. and {Ables}, J.~G. and {Komesaroff}, M.~M.},
        title = "{Detection of change in rotation measure of the Vela pulsar.}",
      journal = {\nat},
         year = 1977,
        month = jan,
       volume = {265},
        pages = {224},
          doi = {10.1038/265224a0},
       adsurl = {https://ui.adsabs.harvard.edu/abs/1977Natur.265..224H}
}

@ARTICLE{Hamilton1977b,
       author = {{Hamilton}, P.~A. and {McCulloch}, P.~M. and {Ables}, J.~G. and {Komesaroff}, M.~M.},
        title = "{Polarization characteristics of southern pulsars - I. 400-MHz observations.}",
      journal = {\mnras},
         year = 1977,
        month = jul,
       volume = {180},
        pages = {1},
          doi = {10.1093/mnras/180.1.1},
       adsurl = {https://ui.adsabs.harvard.edu/abs/1977MNRAS.180....1H}
}

@ARTICLE{Hamilton1985,
       author = {{Hamilton}, P.~A. and {Hall}, P.~J. and {Costa}, M.~E.},
        title = "{Changing parameters along the path to the VELA pulsar.}",
      journal = {\mnras},
         year = 1985,
        month = may,
       volume = {214},
        pages = {5P-8},
          doi = {10.1093/mnras/214.1.5P},
       adsurl = {https://ui.adsabs.harvard.edu/abs/1985MNRAS.214P...5H}
}

@ARTICLE{Rankin1988,
       author = {{Rankin}, Joanna M. and {Campbell}, Donald B. and {Isaacman}, Richard B. and {Payne}, Robert R.},
        title = "{The Crab nebula : secular variations in the Faraday rotation of the pulsar and the great 1974-1975 scattering event.}",
      journal = {\aap},
         year = 1988,
        month = aug,
       volume = {202},
        pages = {166-172},
       adsurl = {https://ui.adsabs.harvard.edu/abs/1988A&A...202..166R}
}

@ARTICLE{vanOmmen1997,
       author = {{van Ommen}, T.~D. and {D'Alessandro}, F. and {Hamilton}, P.~A. and {McCulloch}, P.~M.},
        title = "{Polarimetric observations of southern pulsars at 800 and 950 MHz}",
      journal = {\mnras},
         year = 1997,
        month = may,
       volume = {287},
       number = {2},
        pages = {307-327},
          doi = {10.1093/mnras/287.2.307},
       adsurl = {https://ui.adsabs.harvard.edu/abs/1997MNRAS.287..307V}
}

@ARTICLE{Petroff2013,
       author = {{Petroff}, E. and {Keith}, M.~J. and {Johnston}, S. and {van Straten}, W. and {Shannon}, R.~M.},
        title = "{Dispersion measure variations in a sample of 168 pulsars}",
      journal = {\mnras},
         year = 2013,
        month = oct,
       volume = {435},
       number = {2},
        pages = {1610-1617},
          doi = {10.1093/mnras/stt1401},
archivePrefix = {arXiv},
       eprint = {1307.7221},
 primaryClass = {astro-ph.GA},
       adsurl = {https://ui.adsabs.harvard.edu/abs/2013MNRAS.435.1610P}
}

@INPROCEEDINGS{Keane2015,
       author = {{Keane}, E. and {Bhattacharyya}, B. and {Kramer}, M. and {Stappers}, B. and {Keane}, E.~F. and {Bhattacharyya}, B. and {Kramer}, M. and {Stappers}, B.~W. and {Bates}, S.~D. and {Burgay}, M. and {Chatterjee}, S. and {Champion}, D.~J. and {Eatough}, R.~P. and {Hessels}, J.~W.~T. and {Janssen}, G. and {Lee}, K.~J. and {van Leeuwen}, J. and {Margueron}, J. and {Oertel}, M. and {Possenti}, A. and {Ransom}, S. and {Theureau}, G. and {Torne}, P.},
        title = "{A Cosmic Census of Radio Pulsars with the SKA}",
    booktitle = "{Advancing Astrophysics with the Square Kilometre Array (AASKA14)}",
         year = 2015,
        month = apr,
          eid = {40},
        pages = {40},
          doi = {10.22323/1.215.0040},
archivePrefix = {arXiv},
       eprint = {1501.00056},
 primaryClass = {astro-ph.IM},
       adsurl = {https://ui.adsabs.harvard.edu/abs/2015aska.confE..40K}
}

@ARTICLE{Xue2019,
       author = {{Xue}, Mengyao and {Ord}, S.~M. and {Tremblay}, S.~E. and {Bhat}, N.~D.~R. and {Sobey}, C. and {Meyers}, B.~W. and {McSweeney}, S.~J. and {Swainston}, N.~A.},
        title = "{MWA tied-array processing II: Polarimetric verification and analysis of two bright southern pulsars}",
      journal = {\pasa},
         year = 2019,
        month = jul,
       volume = {36},
          eid = {e025},
        pages = {e025},
          doi = {10.1017/pasa.2019.19},
archivePrefix = {arXiv},
       eprint = {1905.00598},
 primaryClass = {astro-ph.HE},
       adsurl = {https://ui.adsabs.harvard.edu/abs/2019PASA...36...25X}
}

@ARTICLE{Tingay2013,
       author = {{Tingay}, S.~J. and {Goeke}, R. and {Bowman}, J.~D. and {Emrich}, D. and {Ord}, S.~M. and {Mitchell}, D.~A. and {Morales}, M.~F. and {Booler}, T. and {Crosse}, B. and {Wayth}, R.~B. and {Lonsdale}, C.~J. and {Tremblay}, S. and {Pallot}, D. and {Colegate}, T. and {Wicenec}, A. and {Kudryavtseva}, N. and {Arcus}, W. and {Barnes}, D. and {Bernardi}, G. and {Briggs}, F. and {Burns}, S. and {Bunton}, J.~D. and {Cappallo}, R.~J. and {Corey}, B.~E. and {Deshpande}, A. and {Desouza}, L. and {Gaensler}, B.~M. and {Greenhill}, L.~J. and {Hall}, P.~J. and {Hazelton}, B.~J. and {Herne}, D. and {Hewitt}, J.~N. and {Johnston-Hollitt}, M. and {Kaplan}, D.~L. and {Kasper}, J.~C. and {Kincaid}, B.~B. and {Koenig}, R. and {Kratzenberg}, E. and {Lynch}, M.~J. and {Mckinley}, B. and {Mcwhirter}, S.~R. and {Morgan}, E. and {Oberoi}, D. and {Pathikulangara}, J. and {Prabu}, T. and {Remillard}, R.~A. and {Rogers}, A.~E.~E. and {Roshi}, A. and {Salah}, J.~E. and {Sault}, R.~J. and {Udaya-Shankar}, N. and {Schlagenhaufer}, F. and {Srivani}, K.~S. and {Stevens}, J. and {Subrahmanyan}, R. and {Waterson}, M. and {Webster}, R.~L. and {Whitney}, A.~R. and {Williams}, A. and {Williams}, C.~L. and {Wyithe}, J.~S.~B.},
        title = "{The Murchison Widefield Array: The Square Kilometre Array Precursor at Low Radio Frequencies}",
      journal = {\pasa},
         year = 2013,
        month = jan,
       volume = {30},
          eid = {e007},
        pages = {e007},
          doi = {10.1017/pasa.2012.007},
archivePrefix = {arXiv},
       eprint = {1206.6945},
 primaryClass = {astro-ph.IM},
       adsurl = {https://ui.adsabs.harvard.edu/abs/2013PASA...30....7T}
}

@ARTICLE{Wayth2018,
       author = {{Wayth}, Randall B. and {Tingay}, Steven J. and {Trott}, Cathryn M. and {Emrich}, David and {Johnston-Hollitt}, Melanie and {McKinley}, Ben and {Gaensler}, B.~M. and {Beardsley}, A.~P. and {Booler}, T. and {Crosse}, B. and {Franzen}, T.~M.~O. and {Horsley}, L. and {Kaplan}, D.~L. and {Kenney}, D. and {Morales}, M.~F. and {Pallot}, D. and {Sleap}, G. and {Steele}, K. and {Walker}, M. and {Williams}, A. and {Wu}, C. and {Cairns}, Iver. H. and {Filipovic}, M.~D. and {Johnston}, S. and {Murphy}, T. and {Quinn}, P. and {Staveley-Smith}, L. and {Webster}, R. and {Wyithe}, J.~S.~B.},
        title = "{The Phase II Murchison Widefield Array: Design overview}",
      journal = {\pasa},
         year = 2018,
        month = nov,
       volume = {35},
          eid = {e033},
        pages = {e033},
          doi = {10.1017/pasa.2018.37},
archivePrefix = {arXiv},
       eprint = {1809.06466},
 primaryClass = {astro-ph.IM},
       adsurl = {https://ui.adsabs.harvard.edu/abs/2018PASA...35...33W}
}

@ARTICLE{Burn1966,
       author = {{Burn}, B.~J.},
        title = "{On the depolarization of discrete radio sources by Faraday dispersion}",
      journal = {\mnras},
         year = 1966,
        month = jan,
       volume = {133},
        pages = {67},
          doi = {10.1093/mnras/133.1.67},
       adsurl = {https://ui.adsabs.harvard.edu/abs/1966MNRAS.133...67B}
}

@ARTICLE{Brentjens2005,
       author = {{Brentjens}, M.~A. and {de Bruyn}, A.~G.},
        title = "{Faraday rotation measure synthesis}",
      journal = {\aap},
         year = 2005,
        month = oct,
       volume = {441},
       number = {3},
        pages = {1217-1228},
          doi = {10.1051/0004-6361:20052990},
archivePrefix = {arXiv},
       eprint = {astro-ph/0507349},
 primaryClass = {astro-ph},
       adsurl = {https://ui.adsabs.harvard.edu/abs/2005A&A...441.1217B}
}

@ARTICLE{Schnitzeler2009,
       author = {{Schnitzeler}, D.~H.~F.~M. and {Katgert}, P. and {de Bruyn}, A.~G.},
        title = "{WSRT Faraday tomography of the Galactic ISM at {\ensuremath{\lambda}} \raisebox{-0.5ex}\textasciitilde 0.86 m. I. The GEMINI data set at (l, b) = (181{\textdegree}, 20{\textdegree})}",
      journal = {\aap},
         year = 2009,
        month = feb,
       volume = {494},
       number = {2},
        pages = {611-622},
          doi = {10.1051/0004-6361:20078912},
archivePrefix = {arXiv},
       eprint = {0810.4211},
 primaryClass = {astro-ph},
       adsurl = {https://ui.adsabs.harvard.edu/abs/2009A&A...494..611S}
}

@ARTICLE{Heald2009,
       author = {{Heald}, G. and {Braun}, R. and {Edmonds}, R.},
        title = "{The Westerbork SINGS survey. II Polarization, Faraday rotation, and magnetic fields}",
      journal = {\aap},
         year = 2009,
        month = aug,
       volume = {503},
       number = {2},
        pages = {409-435},
          doi = {10.1051/0004-6361/200912240},
archivePrefix = {arXiv},
       eprint = {0905.3995},
 primaryClass = {astro-ph.GA},
       adsurl = {https://ui.adsabs.harvard.edu/abs/2009A&A...503..409H}
}

@ARTICLE{Yan2011,
       author = {{Yan}, W.~M. and {Manchester}, R.~N. and {Hobbs}, G. and {van Straten}, W. and {Reynolds}, J.~E. and {Wang}, N. and {Bailes}, M. and {Bhat}, N.~D.~R. and {Burke-Spolaor}, S. and {Champion}, D.~J. and {Chaudhary}, A. and {Coles}, W.~A. and {Hotan}, A.~W. and {Khoo}, J. and {Oslowski}, S. and {Sarkissian}, J.~M. and {Yardley}, D.~R.~B.},
        title = "{Rotation measure variations for 20 millisecond pulsars}",
      journal = {\apss},
         year = 2011,
        month = oct,
       volume = {335},
       number = {2},
        pages = {485-498},
          doi = {10.1007/s10509-011-0756-0},
archivePrefix = {arXiv},
       eprint = {1105.4213},
 primaryClass = {astro-ph.SR},
       adsurl = {https://ui.adsabs.harvard.edu/abs/2011Ap&SS.335..485Y}
}

@ARTICLE{Backer1993,
       author = {{Backer}, D.~C. and {Hama}, S. and {van Hook}, S. and {Foster}, R.~S.},
        title = "{Temporal Variations of Pulsar Dispersion Measures}",
      journal = {\apj},
         year = 1993,
        month = feb,
       volume = {404},
        pages = {636},
          doi = {10.1086/172317},
       adsurl = {https://ui.adsabs.harvard.edu/abs/1993ApJ...404..636B}
}

@ARTICLE{Hobbs2004,
       author = {{Hobbs}, G. and {Lyne}, A.~G. and {Kramer}, M. and {Martin}, C.~E. and {Jordan}, C.},
        title = "{Long-term timing observations of 374 pulsars}",
      journal = {\mnras},
         year = 2004,
        month = oct,
       volume = {353},
       number = {4},
        pages = {1311-1344},
          doi = {10.1111/j.1365-2966.2004.08157.x},
       adsurl = {https://ui.adsabs.harvard.edu/abs/2004MNRAS.353.1311H}
}

@ARTICLE{Lam2016,
       author = {{Lam}, M.~T. and {Cordes}, J.~M. and {Chatterjee}, S. and {Jones}, M.~L. and {McLaughlin}, M.~A. and {Armstrong}, J.~W.},
        title = "{Systematic and Stochastic Variations in Pulsar Dispersion Measures}",
      journal = {\apj},
         year = 2016,
        month = apr,
       volume = {821},
       number = {1},
          eid = {66},
        pages = {66},
          doi = {10.3847/0004-637X/821/1/66},
archivePrefix = {arXiv},
       eprint = {1512.02203},
 primaryClass = {astro-ph.HE},
       adsurl = {https://ui.adsabs.harvard.edu/abs/2016ApJ...821...66L}
}

@ARTICLE{Orus2005,
       author = {{Or{\'u}s}, R. and {Hern{\'a}ndez-Pajares}, M. and {Juan}, J.~M. and {Sanz}, J.},
        title = "{Improvement of global ionospheric VTEC maps by using kriging interpolation technique}",
      journal = {Journal of Atmospheric and Solar-Terrestrial Physics},
         year = 2005,
        month = nov,
       volume = {67},
       number = {16},
        pages = {1598-1609},
          doi = {10.1016/j.jastp.2005.07.017},
       adsurl = {https://ui.adsabs.harvard.edu/abs/2005JASTP..67.1598O}
}

@ARTICLE{Porayko2019,
       author = {{Porayko}, N.~K. and {Noutsos}, A. and {Tiburzi}, C. and {Verbiest}, J.~P.~W. and {Horneffer}, A. and {K{\"u}nsem{\"o}ller}, J. and {Os{\l}owski}, S. and {Kramer}, M. and {Schnitzeler}, D.~H.~F.~M. and {Anderson}, J.~M. and {Br{\"u}ggen}, M. and {Grie{\ss}meier}, J. -M. and {Hoeft}, M. and {Schwarz}, D.~J. and {Serylak}, M. and {Wucknitz}, O.},
        title = "{Testing the accuracy of the ionospheric Faraday rotation corrections through LOFAR observations of bright northern pulsars}",
      journal = {\mnras},
         year = 2019,
        month = mar,
       volume = {483},
       number = {3},
        pages = {4100-4113},
          doi = {10.1093/mnras/sty3324},
archivePrefix = {arXiv},
       eprint = {1812.01463},
 primaryClass = {astro-ph.IM},
       adsurl = {https://ui.adsabs.harvard.edu/abs/2019MNRAS.483.4100P}
}

@ARTICLE{You2012,
       author = {{You}, X.~P. and {Coles}, W.~A. and {Hobbs}, G.~B. and {Manchester}, R.~N.},
        title = "{Measurement of the electron density and magnetic field of the solar wind using millisecond pulsars}",
      journal = {\mnras},
         year = 2012,
        month = may,
       volume = {422},
       number = {2},
        pages = {1160-1165},
          doi = {10.1111/j.1365-2966.2012.20688.x},
archivePrefix = {arXiv},
       eprint = {1202.2263},
 primaryClass = {astro-ph.SR},
       adsurl = {https://ui.adsabs.harvard.edu/abs/2012MNRAS.422.1160Y}
}

@ARTICLE{HP2009,
       author = {{Hern{\'a}ndez-Pajares}, M. and {Juan}, J.~M. and {Sanz}, J. and {Orus}, R. and {Garcia-Rigo}, A. and {Feltens}, J. and {Komjathy}, A. and {Schaer}, S.~C. and {Krankowski}, A.},
        title = "{The IGS VTEC maps: a reliable source of ionospheric information since 1998}",
      journal = {Journal of Geodesy},
         year = 2009,
        month = mar,
       volume = {83},
       number = {3-4},
        pages = {263-275},
          doi = {10.1007/s00190-008-0266-1},
       adsurl = {https://ui.adsabs.harvard.edu/abs/2009JGeod..83..263H}
}

@ARTICLE{RMextract,
    author = {{Mevius}, M.},
    title = "{RMextract}",
    journal = {Astrophysics Source Code Library},
    year = 2018,
    archivePrefix = "ascl",
    eprint = {1806.024},
}

@ARTICLE{Porayko2023,
       author = {{Porayko}, Nataliya K. and {Mevius}, Maaijke and {Hern{\'a}ndez-Pajares}, Manuel and {Tiburzi}, Caterina and {Olivares Pulido}, German and {Liu}, Qi and {Verbiest}, Joris P.~W. and {K{\"u}nsem{\"o}ller}, J{\"o}rn and {Krishnakumar}, Moochickal Ambalappat and {Bak Nielsen}, Ann-Sofie and {Br{\"u}ggen}, Marcus and {Graffigna}, Victoria and {Dettmar}, Ralf-J{\"u}rgen and {Kramer}, Michael and {Os{\l}owski}, Stefan and {Schwarz}, Dominik J. and {Shaifullah}, Golam M. and {Wucknitz}, Olaf},
        title = "{Validation of global ionospheric models using long-term observations of pulsar Faraday rotation with the LOFAR radio telescope}",
      journal = {Journal of Geodesy},
         year = 2023,
        month = dec,
       volume = {97},
       number = {12},
          eid = {116},
        pages = {116},
          doi = {10.1007/s00190-023-01806-1},
       adsurl = {https://ui.adsabs.harvard.edu/abs/2023JGeod..97..116P}
}

@techreport{IONEX,
    author = {{Schaer}, Stefan},
    title = "{IONEX: The IONosphere Map EXchange Format Version 1.1}",
    institution = "{Astronomical Institute, University of Berne, Switzerland}",
    year = 2015,
    url = {https://gssc.esa.int/wp-content/uploads/2018/07/ionex11.pdf},
}

@misc{WMM,
    author = {{NCEI Geomagnetic Modeling Team} and {British Geological Survey}},
    year = 2019,
    title = "{World Magnetic Model 2020}",
    institution = "{NOAA National Centers for Environmental Information}",
    doi = {10.25921/11v3-da71},
}

@ARTICLE{Johnston2005,
       author = {{Johnston}, Simon and {Hobbs}, G. and {Vigeland}, S. and {Kramer}, M. and {Weisberg}, J.~M. and {Lyne}, A.~G.},
        title = "{Evidence for alignment of the rotation and velocity vectors in pulsars}",
      journal = {\mnras},
         year = 2005,
        month = dec,
       volume = {364},
       number = {4},
        pages = {1397-1412},
          doi = {10.1111/j.1365-2966.2005.09669.x},
archivePrefix = {arXiv},
       eprint = {astro-ph/0510260},
 primaryClass = {astro-ph},
       adsurl = {https://ui.adsabs.harvard.edu/abs/2005MNRAS.364.1397J}
}

@ARTICLE{Kirsten2019,
       author = {{Kirsten}, F. and {Bhat}, N.~D.~R. and {Meyers}, B.~W. and {Macquart}, J. -P. and {Tremblay}, S.~E. and {Ord}, S.~M.},
        title = "{Probing Pulsar Scattering between 120 and 280 MHz with the MWA}",
      journal = {\apj},
         year = 2019,
        month = apr,
       volume = {874},
       number = {2},
          eid = {179},
        pages = {179},
          doi = {10.3847/1538-4357/ab0c05},
archivePrefix = {arXiv},
       eprint = {1903.02087},
 primaryClass = {astro-ph.HE},
       adsurl = {https://ui.adsabs.harvard.edu/abs/2019ApJ...874..179K}
}

@ARTICLE{Lenc2017,
       author = {{Lenc}, E. and {Anderson}, C.~S. and {Barry}, N. and {Bowman}, J.~D. and {Cairns}, I.~H. and {Farnes}, J.~S. and {Gaensler}, B.~M. and {Heald}, G. and {Johnston-Hollitt}, M. and {Kaplan}, D.~L. and {Lynch}, C.~R. and {McCauley}, P.~I. and {Mitchell}, D.~A. and {Morgan}, J. and {Morales}, M.~F. and {Murphy}, Tara and {Offringa}, A.~R. and {Ord}, S.~M. and {Pindor}, B. and {Riseley}, C. and {Sadler}, E.~M. and {Sobey}, C. and {Sokolowski}, M. and {Sullivan}, I.~S. and {O'Sullivan}, S.~P. and {Sun}, X.~H. and {Tremblay}, S.~E. and {Trott}, C.~M. and {Wayth}, R.~B.},
        title = "{The Challenges of Low-Frequency Radio Polarimetry: Lessons from the Murchison Widefield Array}",
      journal = {\pasa},
         year = 2017,
        month = sep,
       volume = {34},
          eid = {e040},
        pages = {e040},
          doi = {10.1017/pasa.2017.36},
archivePrefix = {arXiv},
       eprint = {1708.05799},
 primaryClass = {astro-ph.IM},
       adsurl = {https://ui.adsabs.harvard.edu/abs/2017PASA...34...40L}
}

@ARTICLE{Sobey2021,
       author = {{Sobey}, C. and {Johnston}, S. and {Dai}, S. and {Kerr}, M. and {Manchester}, R.~N. and {Oswald}, L.~S. and {Parthasarathy}, A. and {Shannon}, R.~M. and {Weltevrede}, P.},
        title = "{A polarization census of bright pulsars using the ultrawideband receiver on the Parkes radio telescope}",
      journal = {\mnras},
         year = 2021,
        month = jun,
       volume = {504},
       number = {1},
        pages = {228-247},
          doi = {10.1093/mnras/stab861},
archivePrefix = {arXiv},
       eprint = {2103.13838},
 primaryClass = {astro-ph.HE},
       adsurl = {https://ui.adsabs.harvard.edu/abs/2021MNRAS.504..228S}
}

@ARTICLE{Radhakrishnan1969,
       author = {{Radhakrishnan}, V. and {Cooke}, D.~J.},
        title = "{Magnetic Poles and the Polarization Structure of Pulsar Radiation}",
      journal = {\aplett},
         year = 1969,
        month = jan,
       volume = {3},
        pages = {225},
       adsurl = {https://ui.adsabs.harvard.edu/abs/1969ApL.....3..225R}
}

@ARTICLE{Weltevrede2016,
       author = {{Weltevrede}, P.},
        title = "{Investigation of the bi-drifting subpulses of radio pulsar B1839-04 utilising the open-source data-analysis project PSRSALSA}",
      journal = {\aap},
         year = 2016,
        month = may,
       volume = {590},
          eid = {A109},
        pages = {A109},
          doi = {10.1051/0004-6361/201527950},
archivePrefix = {arXiv},
       eprint = {1605.06413},
 primaryClass = {astro-ph.HE},
       adsurl = {https://ui.adsabs.harvard.edu/abs/2016A&A...590A.109W}
}

@ARTICLE{Noutsos2008,
       author = {{Noutsos}, A. and {Johnston}, S. and {Kramer}, M. and {Karastergiou}, A.},
        title = "{New pulsar rotation measures and the Galactic magnetic field}",
      journal = {\mnras},
         year = 2008,
        month = jun,
       volume = {386},
       number = {4},
        pages = {1881-1896},
          doi = {10.1111/j.1365-2966.2008.13188.x},
archivePrefix = {arXiv},
       eprint = {0803.0677},
 primaryClass = {astro-ph},
       adsurl = {https://ui.adsabs.harvard.edu/abs/2008MNRAS.386.1881N}
}

@ARTICLE{Ashton2019,
       author = {{Ashton}, Gregory and {H{\"u}bner}, Moritz and {Lasky}, Paul D. and {Talbot}, Colm and {Ackley}, Kendall and {Biscoveanu}, Sylvia and {Chu}, Qi and {Divakarla}, Atul and {Easter}, Paul J. and {Goncharov}, Boris and {Hernandez Vivanco}, Francisco and {Harms}, Jan and {Lower}, Marcus E. and {Meadors}, Grant D. and {Melchor}, Denyz and {Payne}, Ethan and {Pitkin}, Matthew D. and {Powell}, Jade and {Sarin}, Nikhil and {Smith}, Rory J.~E. and {Thrane}, Eric},
        title = "{BILBY: A User-friendly Bayesian Inference Library for Gravitational-wave Astronomy}",
      journal = {\apjs},
         year = 2019,
        month = apr,
       volume = {241},
       number = {2},
          eid = {27},
        pages = {27},
          doi = {10.3847/1538-4365/ab06fc},
archivePrefix = {arXiv},
       eprint = {1811.02042},
 primaryClass = {astro-ph.IM},
       adsurl = {https://ui.adsabs.harvard.edu/abs/2019ApJS..241...27A}
}

@ARTICLE{Hobbs2006,
       author = {{Hobbs}, G.~B. and {Edwards}, R.~T. and {Manchester}, R.~N.},
        title = "{TEMPO2, a new pulsar-timing package - I. An overview}",
      journal = {\mnras},
         year = 2006,
        month = jun,
       volume = {369},
       number = {2},
        pages = {655-672},
          doi = {10.1111/j.1365-2966.2006.10302.x},
archivePrefix = {arXiv},
       eprint = {astro-ph/0603381},
 primaryClass = {astro-ph},
       adsurl = {https://ui.adsabs.harvard.edu/abs/2006MNRAS.369..655H}
}

@ARTICLE{Johnston2008,
       author = {{Johnston}, Simon and {Karastergiou}, Aris and {Mitra}, Dipanjan and {Gupta}, Yashwant},
        title = "{Multifrequency integrated profiles of pulsars}",
      journal = {\mnras},
         year = 2008,
        month = jul,
       volume = {388},
       number = {1},
        pages = {261-274},
          doi = {10.1111/j.1365-2966.2008.13379.x},
archivePrefix = {arXiv},
       eprint = {0804.3838},
 primaryClass = {astro-ph},
       adsurl = {https://ui.adsabs.harvard.edu/abs/2008MNRAS.388..261J}
}

@ARTICLE{Johnston2018,
       author = {{Johnston}, Simon and {Kerr}, Matthew},
        title = "{Polarimetry of 600 pulsars from observations at 1.4 GHz with the Parkes radio telescope}",
      journal = {\mnras},
         year = 2018,
        month = mar,
       volume = {474},
       number = {4},
        pages = {4629-4636},
          doi = {10.1093/mnras/stx3095},
archivePrefix = {arXiv},
       eprint = {1711.10092},
 primaryClass = {astro-ph.HE},
       adsurl = {https://ui.adsabs.harvard.edu/abs/2018MNRAS.474.4629J}
}

@ARTICLE{Rookyard2015,
       author = {{Rookyard}, S.~C. and {Weltevrede}, P. and {Johnston}, S.},
        title = "{Constraints on viewing geometries from radio observations of {\ensuremath{\gamma}}-ray-loud pulsars using a novel method}",
      journal = {\mnras},
         year = 2015,
        month = feb,
       volume = {446},
       number = {4},
        pages = {3367-3388},
          doi = {10.1093/mnras/stu2236},
archivePrefix = {arXiv},
       eprint = {1410.3294},
 primaryClass = {astro-ph.HE},
       adsurl = {https://ui.adsabs.harvard.edu/abs/2015MNRAS.446.3367R}
}

@ARTICLE{AstroPy,
       author = {{Astropy Collaboration} and {Price-Whelan}, A.~M. and {Sip{\H{o}}cz}, B.~M. and {G{\"u}nther}, H.~M. and {Lim}, P.~L. and {Crawford}, S.~M. and {Conseil}, S. and {Shupe}, D.~L. and {Craig}, M.~W. and {Dencheva}, N. and {Ginsburg}, A. and {VanderPlas}, J.~T. and {Bradley}, L.~D. and {P{\'e}rez-Su{\'a}rez}, D. and {de Val-Borro}, M. and {Aldcroft}, T.~L. and {Cruz}, K.~L. and {Robitaille}, T.~P. and {Tollerud}, E.~J. and {Ardelean}, C. and {Babej}, T. and {Bach}, Y.~P. and {Bachetti}, M. and {Bakanov}, A.~V. and {Bamford}, S.~P. and {Barentsen}, G. and {Barmby}, P. and {Baumbach}, A. and {Berry}, K.~L. and {Biscani}, F. and {Boquien}, M. and {Bostroem}, K.~A. and {Bouma}, L.~G. and {Brammer}, G.~B. and {Bray}, E.~M. and {Breytenbach}, H. and {Buddelmeijer}, H. and {Burke}, D.~J. and {Calderone}, G. and {Cano Rodr{\'\i}guez}, J.~L. and {Cara}, M. and {Cardoso}, J.~V.~M. and {Cheedella}, S. and {Copin}, Y. and {Corrales}, L. and {Crichton}, D. and {D'Avella}, D. and {Deil}, C. and {Depagne}, {\'E}. and {Dietrich}, J.~P. and {Donath}, A. and {Droettboom}, M. and {Earl}, N. and {Erben}, T. and {Fabbro}, S. and {Ferreira}, L.~A. and {Finethy}, T. and {Fox}, R.~T. and {Garrison}, L.~H. and {Gibbons}, S.~L.~J. and {Goldstein}, D.~A. and {Gommers}, R. and {Greco}, J.~P. and {Greenfield}, P. and {Groener}, A.~M. and {Grollier}, F. and {Hagen}, A. and {Hirst}, P. and {Homeier}, D. and {Horton}, A.~J. and {Hosseinzadeh}, G. and {Hu}, L. and {Hunkeler}, J.~S. and {Ivezi{\'c}}, {\v{Z}}. and {Jain}, A. and {Jenness}, T. and {Kanarek}, G. and {Kendrew}, S. and {Kern}, N.~S. and {Kerzendorf}, W.~E. and {Khvalko}, A. and {King}, J. and {Kirkby}, D. and {Kulkarni}, A.~M. and {Kumar}, A. and {Lee}, A. and {Lenz}, D. and {Littlefair}, S.~P. and {Ma}, Z. and {Macleod}, D.~M. and {Mastropietro}, M. and {McCully}, C. and {Montagnac}, S. and {Morris}, B.~M. and {Mueller}, M. and {Mumford}, S.~J. and {Muna}, D. and {Murphy}, N.~A. and {Nelson}, S. and {Nguyen}, G.~H. and {Ninan}, J.~P. and {N{\"o}the}, M. and {Ogaz}, S. and {Oh}, S. and {Parejko}, J.~K. and {Parley}, N. and {Pascual}, S. and {Patil}, R. and {Patil}, A.~A. and {Plunkett}, A.~L. and {Prochaska}, J.~X. and {Rastogi}, T. and {Reddy Janga}, V. and {Sabater}, J. and {Sakurikar}, P. and {Seifert}, M. and {Sherbert}, L.~E. and {Sherwood-Taylor}, H. and {Shih}, A.~Y. and {Sick}, J. and {Silbiger}, M.~T. and {Singanamalla}, S. and {Singer}, L.~P. and {Sladen}, P.~H. and {Sooley}, K.~A. and {Sornarajah}, S. and {Streicher}, O. and {Teuben}, P. and {Thomas}, S.~W. and {Tremblay}, G.~R. and {Turner}, J.~E.~H. and {Terr{\'o}n}, V. and {van Kerkwijk}, M.~H. and {de la Vega}, A. and {Watkins}, L.~L. and {Weaver}, B.~A. and {Whitmore}, J.~B. and {Woillez}, J. and {Zabalza}, V. and {Astropy Contributors}},
        title = "{The Astropy Project: Building an Open-science Project and Status of the v2.0 Core Package}",
      journal = {\aj},
         year = 2018,
        month = sep,
       volume = {156},
       number = {3},
          eid = {123},
        pages = {123},
          doi = {10.3847/1538-3881/aabc4f},
archivePrefix = {arXiv},
       eprint = {1801.02634},
 primaryClass = {astro-ph.IM},
       adsurl = {https://ui.adsabs.harvard.edu/abs/2018AJ....156..123A}
}

@ARTICLE{Matplotlib,
    author={Hunter, John D.},
    journal={CSE},
    title={Matplotlib: A 2D Graphics Environment},
    year={2007},
    volume={9},
    number={3},
    pages={90-95},
    doi={10.1109/MCSE.2007.55}
}

@ARTICLE{NumPy,
       author = {{Harris}, Charles R. and {Millman}, K. Jarrod and {van der Walt}, St{\'e}fan J. and {Gommers}, Ralf and {Virtanen}, Pauli and {Cournapeau}, David and {Wieser}, Eric and {Taylor}, Julian and {Berg}, Sebastian and {Smith}, Nathaniel J. and {Kern}, Robert and {Picus}, Matti and {Hoyer}, Stephan and {van Kerkwijk}, Marten H. and {Brett}, Matthew and {Haldane}, Allan and {del R{\'\i}o}, Jaime Fern{\'a}ndez and {Wiebe}, Mark and {Peterson}, Pearu and {G{\'e}rard-Marchant}, Pierre and {Sheppard}, Kevin and {Reddy}, Tyler and {Weckesser}, Warren and {Abbasi}, Hameer and {Gohlke}, Christoph and {Oliphant}, Travis E.},
        title = "{Array programming with NumPy}",
      journal = {\nat},
         year = 2020,
        month = sep,
       volume = {585},
       number = {7825},
        pages = {357-362},
          doi = {10.1038/s41586-020-2649-2},
archivePrefix = {arXiv},
       eprint = {2006.10256},
 primaryClass = {cs.MS},
       adsurl = {https://ui.adsabs.harvard.edu/abs/2020Natur.585..357H}
}

@article{Huber1964,
    title = {{Robust Estimation of a Location Parameter}},
    year = {1964},
    journal = {The Annals of Mathematical Statistics},
    author = {Huber, Peter J.},
    number = {1},
    month = {3},
    pages = {73--101},
    volume = {35},
    url = {http://projecteuclid.org/euclid.aoms/1177703732},
    doi = {10.1214/aoms/1177703732},
    issn = {0003-4851}
}

@ARTICLE{Manchester2005,
       author = {{Manchester}, R.~N. and {Hobbs}, G.~B. and {Teoh}, A. and {Hobbs}, M.},
        title = "{The Australia Telescope National Facility Pulsar Catalogue}",
      journal = {\aj},
         year = 2005,
        month = apr,
       volume = {129},
       number = {4},
        pages = {1993-2006},
          doi = {10.1086/428488},
archivePrefix = {arXiv},
       eprint = {astro-ph/0412641},
 primaryClass = {astro-ph},
       adsurl = {https://ui.adsabs.harvard.edu/abs/2005AJ....129.1993M}
}

@ARTICLE{Dodson2003,
       author = {{Dodson}, R. and {Legge}, D. and {Reynolds}, J.~E. and {McCulloch}, P.~M.},
        title = "{The Vela Pulsar's Proper Motion and Parallax Derived from VLBI Observations}",
      journal = {\apj},
         year = 2003,
        month = oct,
       volume = {596},
       number = {2},
        pages = {1137-1141},
          doi = {10.1086/378089},
archivePrefix = {arXiv},
       eprint = {astro-ph/0302374},
 primaryClass = {astro-ph},
       adsurl = {https://ui.adsabs.harvard.edu/abs/2003ApJ...596.1137D}
}

@ARTICLE{Posselt2023,
       author = {{Posselt}, B. and {Karastergiou}, A. and {Johnston}, S. and {Parthasarathy}, A. and {Oswald}, L.~S. and {Main}, R.~A. and {Basu}, A. and {Keith}, M.~J. and {Song}, X. and {Weltevrede}, P. and {Tiburzi}, C. and {Bailes}, M. and {Buchner}, S. and {Geyer}, M. and {Kramer}, M. and {Spiewak}, R. and {Krishnan}, V. Venkatraman},
        title = "{The Thousand Pulsar Array program on MeerKAT - IX. The time-averaged properties of the observed pulsar population}",
      journal = {\mnras},
         year = 2023,
        month = apr,
       volume = {520},
       number = {3},
        pages = {4582-4600},
          doi = {10.1093/mnras/stac3383},
archivePrefix = {arXiv},
       eprint = {2211.11849},
 primaryClass = {astro-ph.HE},
       adsurl = {https://ui.adsabs.harvard.edu/abs/2023MNRAS.520.4582P}
}

@ARTICLE{Lyne1994,
       author = {{Lyne}, A.~G. and {Lorimer}, D.~R.},
        title = "{High birth velocities of radio pulsars}",
      journal = {\nat},
         year = 1994,
        month = may,
       volume = {369},
       number = {6476},
        pages = {127-129},
          doi = {10.1038/369127a0},
       adsurl = {https://ui.adsabs.harvard.edu/abs/1994Natur.369..127L}
}

@ARTICLE{Osterbrock1957,
       author = {{Osterbrock}, Donald E.},
        title = "{Electron Densities in the Filaments of the Crab Nebula}",
      journal = {\pasp},
         year = 1957,
        month = jun,
       volume = {69},
       number = {408},
        pages = {227},
          doi = {10.1086/127053},
       adsurl = {https://ui.adsabs.harvard.edu/abs/1957PASP...69..227O}
}

@ARTICLE{Woltjer1958,
       author = {{Woltjer}, L.},
        title = "{The Crab nebula}",
      journal = {\bain},
         year = 1958,
        month = jan,
       volume = {14},
        pages = {39},
       adsurl = {https://ui.adsabs.harvard.edu/abs/1958BAN....14...39W}
}

@article{corner,
      doi = {10.21105/joss.00024},
      url = {https://doi.org/10.21105/joss.00024},
      year  = {2016},
      month = {jun},
      publisher = {The Open Journal},
      volume = {1},
      number = {2},
      pages = {24},
      author = {Daniel Foreman-Mackey},
      title = {corner.py: Scatterplot matrices in Python},
      journal = {The Journal of Open Source Software}
    }

@ARTICLE{Jankowski2018,
       author = {{Jankowski}, F. and {van Straten}, W. and {Keane}, E.~F. and {Bailes}, M. and {Barr}, E.~D. and {Johnston}, S. and {Kerr}, M.},
        title = "{Spectral properties of 441 radio pulsars}",
      journal = {\mnras},
         year = 2018,
        month = feb,
       volume = {473},
       number = {4},
        pages = {4436-4458},
          doi = {10.1093/mnras/stx2476},
archivePrefix = {arXiv},
       eprint = {1709.08864},
 primaryClass = {astro-ph.HE},
       adsurl = {https://ui.adsabs.harvard.edu/abs/2018MNRAS.473.4436J}
}

@ARTICLE{Speagle2020,
       author = {{Speagle}, Joshua S.},
        title = "{DYNESTY: a dynamic nested sampling package for estimating Bayesian posteriors and evidences}",
      journal = {\mnras},
         year = 2020,
        month = apr,
       volume = {493},
       number = {3},
        pages = {3132-3158},
          doi = {10.1093/mnras/staa278},
archivePrefix = {arXiv},
       eprint = {1904.02180},
 primaryClass = {astro-ph.IM},
       adsurl = {https://ui.adsabs.harvard.edu/abs/2020MNRAS.493.3132S}
}

@ARTICLE{Jordan2017,
       author = {{Jordan}, C.~H. and {Murray}, S. and {Trott}, C.~M. and {Wayth}, R.~B. and {Mitchell}, D.~A. and {Rahimi}, M. and {Pindor}, B. and {Procopio}, P. and {Morgan}, J.},
        title = "{Characterization of the ionosphere above the Murchison Radio Observatory using the Murchison Widefield Array}",
      journal = {\mnras},
         year = 2017,
        month = nov,
       volume = {471},
       number = {4},
        pages = {3974-3987},
          doi = {10.1093/mnras/stx1797},
archivePrefix = {arXiv},
       eprint = {1707.04978},
 primaryClass = {astro-ph.IM},
       adsurl = {https://ui.adsabs.harvard.edu/abs/2017MNRAS.471.3974J}
}

@ARTICLE{Keith2024,
       author = {{Keith}, M.~J. and {Johnston}, S. and {Karastergiou}, A. and {Weltevrede}, P. and {Lower}, M.~E. and {Basu}, A. and {Posselt}, B. and {Oswald}, L.~S. and {Parthasarathy}, A. and {Cameron}, A.~D. and {Serylak}, M. and {Buchner}, S.},
        title = "{The Thousand-Pulsar-Array programme on MeerKAT - XIII. Timing, flux density, rotation measure, and dispersion measure time series of 597 pulsars}",
      journal = {\mnras},
         year = 2024,
        month = may,
       volume = {530},
       number = {2},
        pages = {1581-1591},
          doi = {10.1093/mnras/stae937},
archivePrefix = {arXiv},
       eprint = {2404.02051},
 primaryClass = {astro-ph.HE},
       adsurl = {https://ui.adsabs.harvard.edu/abs/2024MNRAS.530.1581K}
}

@ARTICLE{Heiles1989,
       author = {{Heiles}, Carl},
        title = "{Magnetic Fields, Pressures, and Thermally Unstable Gas in Prominent H i Shells}",
      journal = {\apj},
         year = 1989,
        month = jan,
       volume = {336},
        pages = {808},
          doi = {10.1086/167051},
       adsurl = {https://ui.adsabs.harvard.edu/abs/1989ApJ...336..808H}
}

@ARTICLE{Beck2005,
       author = {{Beck}, R. and {Krause}, M.},
        title = "{Revised equipartition and minimum energy formula for magnetic field strength estimates from radio synchrotron observations}",
      journal = {Astronomische Nachrichten},
         year = 2005,
        month = jul,
       volume = {326},
       number = {6},
        pages = {414-427},
          doi = {10.1002/asna.200510366},
archivePrefix = {arXiv},
       eprint = {astro-ph/0507367},
 primaryClass = {astro-ph},
       adsurl = {https://ui.adsabs.harvard.edu/abs/2005AN....326..414B}
}

@ARTICLE{Arbutina2012,
       author = {{Arbutina}, B. and {Uro{\v{s}}evi{\'c}}, D. and {Andjeli{\'c}}, M.~M. and {Pavlovi{\'c}}, M.~Z. and {Vukoti{\'c}}, B.},
        title = "{Modified Equipartition Calculation for Supernova Remnants}",
      journal = {\apj},
         year = 2012,
        month = feb,
       volume = {746},
       number = {1},
          eid = {79},
        pages = {79},
          doi = {10.1088/0004-637X/746/1/79},
archivePrefix = {arXiv},
       eprint = {1111.5465},
 primaryClass = {astro-ph.HE},
       adsurl = {https://ui.adsabs.harvard.edu/abs/2012ApJ...746...79A}
}

@ARTICLE{Urosevic2018,
       author = {{Uro{\v{s}}evi{\'c}}, Dejan and {Pavlovi{\'c}}, Marko Z. and {Arbutina}, Bojan},
        title = "{On the Foundation of Equipartition in Supernova Remnants}",
      journal = {\apj},
         year = 2018,
        month = mar,
       volume = {855},
       number = {1},
          eid = {59},
        pages = {59},
          doi = {10.3847/1538-4357/aaac2d},
archivePrefix = {arXiv},
       eprint = {1801.10422},
 primaryClass = {astro-ph.HE},
       adsurl = {https://ui.adsabs.harvard.edu/abs/2018ApJ...855...59U}
}

@ARTICLE{Heiles1976,
       author = {{Heiles}, C.},
        title = "{The interstellar magnetic field}",
      journal = {\araa},
         year = 1976,
        month = jan,
       volume = {14},
        pages = {1-22},
          doi = {10.1146/annurev.aa.14.090176.000245},
       adsurl = {https://ui.adsabs.harvard.edu/abs/1976ARA&A..14....1H}
}

@ARTICLE{Beck2003,
       author = {{Beck}, R. and {Shukurov}, A. and {Sokoloff}, D. and {Wielebinski}, R.},
        title = "{Systematic bias in interstellar magnetic field estimates}",
      journal = {\aap},
         year = 2003,
        month = nov,
       volume = {411},
        pages = {99-107},
          doi = {10.1051/0004-6361:20031101},
archivePrefix = {arXiv},
       eprint = {astro-ph/0307330},
 primaryClass = {astro-ph},
       adsurl = {https://ui.adsabs.harvard.edu/abs/2003A&A...411...99B}
}

@ARTICLE{Manchester1972,
       author = {{Manchester}, R.~N.},
        title = "{Pulsar Rotation and Dispersion Measures and the Galactic Magnetic Field.}",
      journal = {\apj},
         year = 1972,
        month = feb,
       volume = {172},
        pages = {43},
          doi = {10.1086/151326},
       adsurl = {https://ui.adsabs.harvard.edu/abs/1972ApJ...172...43M}
}

@ARTICLE{Manchester1974,
       author = {{Manchester}, R.~N.},
        title = "{Structure of the Local Galactic Magnetic Field}",
      journal = {\apj},
         year = 1974,
        month = mar,
       volume = {188},
        pages = {637-644},
          doi = {10.1086/152757},
       adsurl = {https://ui.adsabs.harvard.edu/abs/1974ApJ...188..637M}
}

@ARTICLE{Han2006,
       author = {{Han}, J.~L. and {Manchester}, R.~N. and {Lyne}, A.~G. and {Qiao}, G.~J. and {van Straten}, W.},
        title = "{Pulsar Rotation Measures and the Large-Scale Structure of the Galactic Magnetic Field}",
      journal = {\apj},
         year = 2006,
        month = may,
       volume = {642},
       number = {2},
        pages = {868-881},
          doi = {10.1086/501444},
archivePrefix = {arXiv},
       eprint = {astro-ph/0601357},
 primaryClass = {astro-ph},
       adsurl = {https://ui.adsabs.harvard.edu/abs/2006ApJ...642..868H}
}

@ARTICLE{Sobey2019,
       author = {{Sobey}, C. and {Bilous}, A.~V. and {Grie{\ss}meier}, J. -M. and {Hessels}, J.~W.~T. and {Karastergiou}, A. and {Keane}, E.~F. and {Kondratiev}, V.~I. and {Kramer}, M. and {Michilli}, D. and {Noutsos}, A. and {Pilia}, M. and {Polzin}, E.~J. and {Stappers}, B.~W. and {Tan}, C.~M. and {van Leeuwen}, J. and {Verbiest}, J.~P.~W. and {Weltevrede}, P. and {Heald}, G. and {Alves}, M.~I.~R. and {Carretti}, E. and {En{\ss}lin}, T. and {Haverkorn}, M. and {Iacobelli}, M. and {Reich}, W. and {Van Eck}, C.},
        title = "{Low-frequency Faraday rotation measures towards pulsars using LOFAR: probing the 3D Galactic halo magnetic field}",
      journal = {\mnras},
         year = 2019,
        month = apr,
       volume = {484},
       number = {3},
        pages = {3646-3664},
          doi = {10.1093/mnras/stz214},
archivePrefix = {arXiv},
       eprint = {1901.07738},
 primaryClass = {astro-ph.GA},
       adsurl = {https://ui.adsabs.harvard.edu/abs/2019MNRAS.484.3646S}
}

@ARTICLE{Howard2016,
       author = {{Howard}, T.~A. and {Stovall}, K. and {Dowell}, J. and {Taylor}, G.~B. and {White}, S.~M.},
        title = "{Measuring the Magnetic Field of Coronal Mass Ejections Near the Sun Using Pulsars}",
      journal = {\apj},
         year = 2016,
        month = nov,
       volume = {831},
       number = {2},
          eid = {208},
        pages = {208},
          doi = {10.3847/0004-637X/831/2/208},
       adsurl = {https://ui.adsabs.harvard.edu/abs/2016ApJ...831..208H}
}

@ARTICLE{Bhat1998,
       author = {{Bhat}, N.~D. Ramesh and {Gupta}, Yashwant and {Rao}, A. Pramesh},
        title = "{Pulsar Scintillation and the Local Bubble}",
      journal = {\apj},
         year = 1998,
        month = jun,
       volume = {500},
       number = {1},
        pages = {262-279},
          doi = {10.1086/305715},
archivePrefix = {arXiv},
       eprint = {astro-ph/9802203},
 primaryClass = {astro-ph},
       adsurl = {https://ui.adsabs.harvard.edu/abs/1998ApJ...500..262B}
}

@ARTICLE{Seta2022,
       author = {{Seta}, Amit and {Federrath}, Christoph},
        title = "{Turbulent dynamo in the two-phase interstellar medium}",
      journal = {\mnras},
         year = 2022,
        month = jul,
       volume = {514},
       number = {1},
        pages = {957-976},
          doi = {10.1093/mnras/stac1400},
archivePrefix = {arXiv},
       eprint = {2202.08324},
 primaryClass = {astro-ph.GA},
       adsurl = {https://ui.adsabs.harvard.edu/abs/2022MNRAS.514..957S}
}

@ARTICLE{Reardon2020,
       author = {{Reardon}, Daniel J. and {Coles}, William A. and {Bailes}, Matthew and {Bhat}, N.~D. Ramesh and {Dai}, Shi and {Hobbs}, George B. and {Kerr}, Matthew and {Manchester}, Richard N. and {Os{\l}owski}, Stefan and {Parthasarathy}, Aditya and {Russell}, Christopher J. and {Shannon}, Ryan M. and {Spiewak}, Ren{\'e}e and {Toomey}, Lawrence and {Tuntsov}, Artem V. and {van Straten}, Willem and {Walker}, Mark A. and {Wang}, Jingbo and {Zhang}, Lei and {Zhu}, Xing-Jiang},
        title = "{Precision Orbital Dynamics from Interstellar Scintillation Arcs for PSR J0437-4715}",
      journal = {\apj},
         year = 2020,
        month = dec,
       volume = {904},
       number = {2},
          eid = {104},
        pages = {104},
          doi = {10.3847/1538-4357/abbd40},
archivePrefix = {arXiv},
       eprint = {2009.12757},
 primaryClass = {astro-ph.HE},
       adsurl = {https://ui.adsabs.harvard.edu/abs/2020ApJ...904..104R}
}

@ARTICLE{Dong2018,
       author = {{Dong}, Lingyi and {Petropoulou}, Maria and {Giannios}, Dimitrios},
        title = "{Extreme scattering events from axisymmetric plasma lenses}",
      journal = {\mnras},
         year = 2018,
        month = dec,
       volume = {481},
       number = {2},
        pages = {2685-2693},
          doi = {10.1093/mnras/sty2427},
archivePrefix = {arXiv},
       eprint = {1809.00005},
 primaryClass = {astro-ph.GA},
       adsurl = {https://ui.adsabs.harvard.edu/abs/2018MNRAS.481.2685D}
}

@ARTICLE{Bhat2004,
       author = {{Bhat}, N.~D. Ramesh and {Cordes}, James M. and {Camilo}, Fernando and {Nice}, David J. and {Lorimer}, Duncan R.},
        title = "{Multifrequency Observations of Radio Pulse Broadening and Constraints on Interstellar Electron Density Microstructure}",
      journal = {\apj},
         year = 2004,
        month = apr,
       volume = {605},
       number = {2},
        pages = {759-783},
          doi = {10.1086/382680},
archivePrefix = {arXiv},
       eprint = {astro-ph/0401067},
 primaryClass = {astro-ph},
       adsurl = {https://ui.adsabs.harvard.edu/abs/2004ApJ...605..759B}
}

@ARTICLE{Geyer2017,
       author = {{Geyer}, M. and {Karastergiou}, A. and {Kondratiev}, V.~I. and {Zagkouris}, K. and {Kramer}, M. and {Stappers}, B.~W. and {Grie{\ss}meier}, J. -M. and {Hessels}, J.~W.~T. and {Michilli}, D. and {Pilia}, M. and {Sobey}, C.},
        title = "{Scattering analysis of LOFAR pulsar observations}",
      journal = {\mnras},
         year = 2017,
        month = sep,
       volume = {470},
       number = {3},
        pages = {2659-2679},
          doi = {10.1093/mnras/stx1151},
archivePrefix = {arXiv},
       eprint = {1706.04205},
 primaryClass = {astro-ph.HE},
       adsurl = {https://ui.adsabs.harvard.edu/abs/2017MNRAS.470.2659G}
}

@ARTICLE{Fiedler1987,
       author = {{Fiedler}, R.~L. and {Dennison}, B. and {Johnston}, K.~J. and {Hewish}, A.},
        title = "{Extreme scattering events caused by compact structures in the interstellar medium}",
      journal = {\nat},
         year = 1987,
        month = apr,
       volume = {326},
       number = {6114},
        pages = {675-678},
          doi = {10.1038/326675a0},
       adsurl = {https://ui.adsabs.harvard.edu/abs/1987Natur.326..675F}
}

@ARTICLE{Tiburzi2021,
       author = {{Tiburzi}, C. and {Shaifullah}, G.~M. and {Bassa}, C.~G. and {Zucca}, P. and {Verbiest}, J.~P.~W. and {Porayko}, N.~K. and {van der Wateren}, E. and {Fallows}, R.~A. and {Main}, R.~A. and {Janssen}, G.~H. and {Anderson}, J.~M. and {Bak Nielsen}, A. -S. and {Donner}, J.~Y. and {Keane}, E.~F. and {K{\"u}nsem{\"o}ller}, J. and {Os{\l}owski}, S. and {Grie{\ss}meier}, J. -M. and {Serylak}, M. and {Br{\"u}ggen}, M. and {Ciardi}, B. and {Dettmar}, R. -J. and {Hoeft}, M. and {Kramer}, M. and {Mann}, G. and {Vocks}, C.},
        title = "{The impact of solar wind variability on pulsar timing}",
      journal = {\aap},
         year = 2021,
        month = mar,
       volume = {647},
          eid = {A84},
        pages = {A84},
          doi = {10.1051/0004-6361/202039846},
archivePrefix = {arXiv},
       eprint = {2012.11726},
 primaryClass = {astro-ph.HE},
       adsurl = {https://ui.adsabs.harvard.edu/abs/2021A&A...647A..84T}
}

@ARTICLE{Reardon2023,
       author = {{Reardon}, Daniel J. and {Zic}, Andrew and {Shannon}, Ryan M. and {Di Marco}, Valentina and {Hobbs}, George B. and {Kapur}, Agastya and {Lower}, Marcus E. and {Mandow}, Rami and {Middleton}, Hannah and {Miles}, Matthew T. and {Rogers}, Axl F. and {Askew}, Jacob and {Bailes}, Matthew and {Bhat}, N.~D. Ramesh and {Cameron}, Andrew and {Kerr}, Matthew and {Kulkarni}, Atharva and {Manchester}, Richard N. and {Nathan}, Rowina S. and {Russell}, Christopher J. and {Os{\l}owski}, Stefan and {Zhu}, Xing-Jiang},
        title = "{The Gravitational-wave Background Null Hypothesis: Characterizing Noise in Millisecond Pulsar Arrival Times with the Parkes Pulsar Timing Array}",
      journal = {\apjl},
         year = 2023,
        month = jul,
       volume = {951},
       number = {1},
          eid = {L7},
        pages = {L7},
          doi = {10.3847/2041-8213/acdd03},
archivePrefix = {arXiv},
       eprint = {2306.16229},
 primaryClass = {astro-ph.HE},
       adsurl = {https://ui.adsabs.harvard.edu/abs/2023ApJ...951L...7R}
}

@ARTICLE{Agazie2023,
       author = {{Agazie}, Gabriella and {Anumarlapudi}, Akash and {Archibald}, Anne M. and {Arzoumanian}, Zaven and {Baker}, Paul T. and {B{\'e}csy}, Bence and {Blecha}, Laura and {Brazier}, Adam and {Brook}, Paul R. and {Burke-Spolaor}, Sarah and {Charisi}, Maria and {Chatterjee}, Shami and {Cohen}, Tyler and {Cordes}, James M. and {Cornish}, Neil J. and {Crawford}, Fronefield and {Cromartie}, H. Thankful and {Crowter}, Kathryn and {Decesar}, Megan E. and {Demorest}, Paul B. and {Dolch}, Timothy and {Drachler}, Brendan and {Ferrara}, Elizabeth C. and {Fiore}, William and {Fonseca}, Emmanuel and {Freedman}, Gabriel E. and {Garver-Daniels}, Nate and {Gentile}, Peter A. and {Glaser}, Joseph and {Good}, Deborah C. and {Guertin}, Lydia and {G{\"u}ltekin}, Kayhan and {Hazboun}, Jeffrey S. and {Jennings}, Ross J. and {Johnson}, Aaron D. and {Jones}, Megan L. and {Kaiser}, Andrew R. and {Kaplan}, David L. and {Kelley}, Luke Zoltan and {Kerr}, Matthew and {Key}, Joey S. and {Laal}, Nima and {Lam}, Michael T. and {Lamb}, William G. and {Lazio}, T. Joseph W. and {Lewandowska}, Natalia and {Liu}, Tingting and {Lorimer}, Duncan R. and {Luo}, Jing and {Lynch}, Ryan S. and {Ma}, Chung-Pei and {Madison}, Dustin R. and {McEwen}, Alexander and {McKee}, James W. and {McLaughlin}, Maura A. and {McMann}, Natasha and {Meyers}, Bradley W. and {Mingarelli}, Chiara M.~F. and {Mitridate}, Andrea and {Ng}, Cherry and {Nice}, David J. and {Ocker}, Stella Koch and {Olum}, Ken D. and {Pennucci}, Timothy T. and {Perera}, Benetge B.~P. and {Pol}, Nihan S. and {Radovan}, Henri A. and {Ransom}, Scott M. and {Ray}, Paul S. and {Romano}, Joseph D. and {Sardesai}, Shashwat C. and {Schmiedekamp}, Ann and {Schmiedekamp}, Carl and {Schmitz}, Kai and {Shapiro-Albert}, Brent J. and {Siemens}, Xavier and {Simon}, Joseph and {Siwek}, Magdalena S. and {Stairs}, Ingrid H. and {Stinebring}, Daniel R. and {Stovall}, Kevin and {Susobhanan}, Abhimanyu and {Swiggum}, Joseph K. and {Taylor}, Stephen R. and {Turner}, Jacob E. and {Unal}, Caner and {Vallisneri}, Michele and {Vigeland}, Sarah J. and {Wahl}, Haley M. and {Witt}, Caitlin A. and {Young}, Olivia and {Nanograv Collaboration}},
        title = "{The NANOGrav 15 yr Data Set: Detector Characterization and Noise Budget}",
      journal = {\apjl},
         year = 2023,
        month = jul,
       volume = {951},
       number = {1},
          eid = {L10},
        pages = {L10},
          doi = {10.3847/2041-8213/acda88},
archivePrefix = {arXiv},
       eprint = {2306.16218},
 primaryClass = {astro-ph.HE},
       adsurl = {https://ui.adsabs.harvard.edu/abs/2023ApJ...951L..10A}
}

@ARTICLE{Clegg1998,
       author = {{Clegg}, Andrew W. and {Fey}, Alan L. and {Lazio}, T. Joseph W.},
        title = "{The Gaussian Plasma Lens in Astrophysics: Refraction}",
      journal = {\apj},
         year = 1998,
        month = mar,
       volume = {496},
       number = {1},
        pages = {253-266},
          doi = {10.1086/305344},
archivePrefix = {arXiv},
       eprint = {astro-ph/9709249},
 primaryClass = {astro-ph},
       adsurl = {https://ui.adsabs.harvard.edu/abs/1998ApJ...496..253C}
}

@ARTICLE{Romani1987,
       author = {{Romani}, Roger W. and {Blandford}, Roger D. and {Cordes}, James M.},
        title = "{Radio caustics from localized interstellar medium plasma structures}",
      journal = {\nat},
         year = 1987,
        month = jul,
       volume = {328},
       number = {6128},
        pages = {324-326},
          doi = {10.1038/328324a0},
       adsurl = {https://ui.adsabs.harvard.edu/abs/1987Natur.328..324R}
}

\appendix

\section{Model fitting methodology}\label{app:model}
We consider two models for the RM and DM as a function of time.
The first is a simple (single-segment) linear model,
\begin{equation}
    f_1(t;m,c) = mt+c,
\end{equation}
where $t$ is the time, $m$ is the gradient, and $c$ is the vertical intercept.
The second is a piecewise (two-segment) linear model,
\begin{equation}
    f_2(t;m_1,m_2,t_b,c) = m_2t + (m_1 - m_2)t_b + c,
\end{equation}
where $m_1$ and $m_2$ are the gradients of the first and second segment, and $t_b$ is the break point between the segments.

Since the data set contains measurements from different telescopes at different observing frequencies, systematic errors will be inconsistent between publications.
To address an analogous issue for pulsar flux density spectra, \citet{Jankowski2018} defined a robust log-likelihood which uses the Huber loss function to reduce the sensitivity to outlier data points \citep{Huber1964}.
The robust log-likelihood is defined as
\begin{equation}
    \log\mathcal{L}(y_i;t_i,\Theta) = -\sum_{i=1}^N
    \begin{cases}
    \frac{1}{2}\left(\frac{y_i-f(t_i;\Theta)}{\sigma_{y,i}}\right)^2 & \mathrm{if}\:\left|\frac{y_i-f(t_i;\Theta)}{\sigma_{y,i}}\right| < k \\[5pt]
    k\left|\frac{y_i-f(t_i;\Theta)}{\sigma_{y,i}}\right|-\frac{1}{2}k^2 & \mathrm{otherwise}
    \end{cases},
\end{equation}
where $f(t_i;\Theta)$ is the model, $\Theta$ are the model parameters, $(t_i, y_i)$ are the data, $\sigma_{y,i}$ are the uncertainties on $y_i$, and $k$ is a parameter defining the point at which outlier data points are penalised.
Following \citet{Huber1964}, we used $k=1.345$.

We jointly fit the RM and DM datasets by combining the likelihoods, forming a total likelihood
\begin{equation}
    \log\mathcal{L} = \log\mathcal{L}_\mathrm{RM} + \log\mathcal{L}_\mathrm{DM}.
\end{equation}
For the piecewise linear model, $\mathcal{L}_\mathrm{RM}$ and $\mathcal{L}_\mathrm{DM}$ both depend on the break-point parameter $t_b$, so the RM and DM models are covariant.

We used Bayesian inference to estimate the model parameters and their uncertainties.
For the vertical intercepts and the break point, the priors were set to be uniform within the span of the dataset, and zero elsewhere.
For the gradients $\mathrm{dRM}/\mathrm{d}t$ and $\mathrm{dDM}/\mathrm{d}t$, the priors were chosen to be uniform in the ranges $[-5,5]$\,\rmuyr{} and $[-0.5,0.5]$\,\dmuyr, respectively.
We used the \texttt{Dynesty} dynamic nested sampler \citep{Speagle2020} integrated within \texttt{Bilby} \citep{Ashton2019} to simultaneously estimate the Bayesian evidence and the posterior probability distribution.
We then tested the sensitivity of the calculated evidence to the chosen priors by varying the size of the prior space.

\begin{figure}
    \centering
    \includegraphics[width=\linewidth]{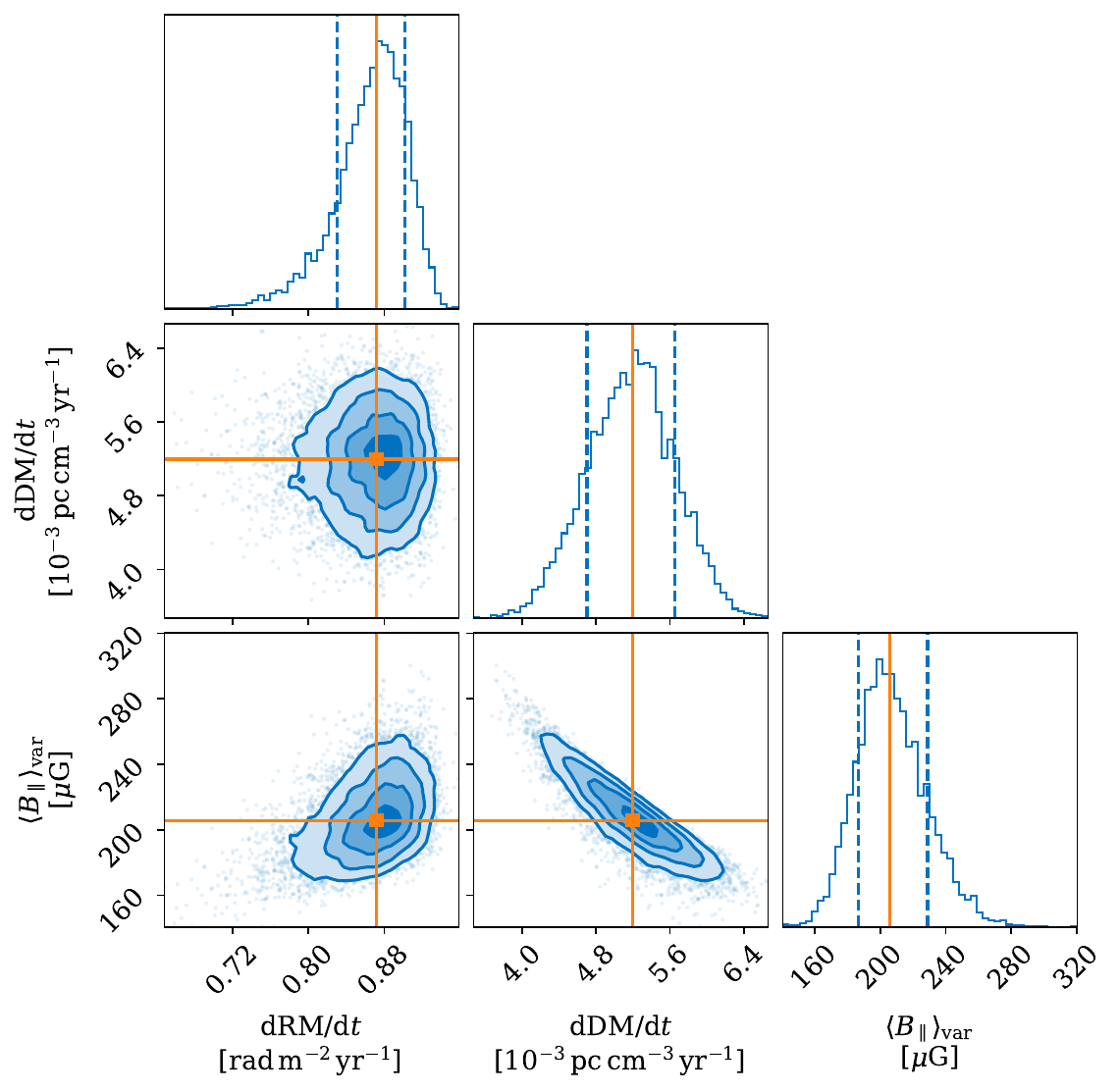}
    \caption{Posterior distributions for the simple linear model fit. The orange lines indicate the median and the blue dashed lines indicate the 16\% and 84\% percentiles of 1D posterior distributions. The contours indicate the 11.8\%, 39.3\%, 67.5\%, and 86.4\% percentiles of the 2D posterior distributions.}
    \label{fig:pos_lin}
\end{figure}

\begin{figure*}
    \centering
    \includegraphics[width=\linewidth]{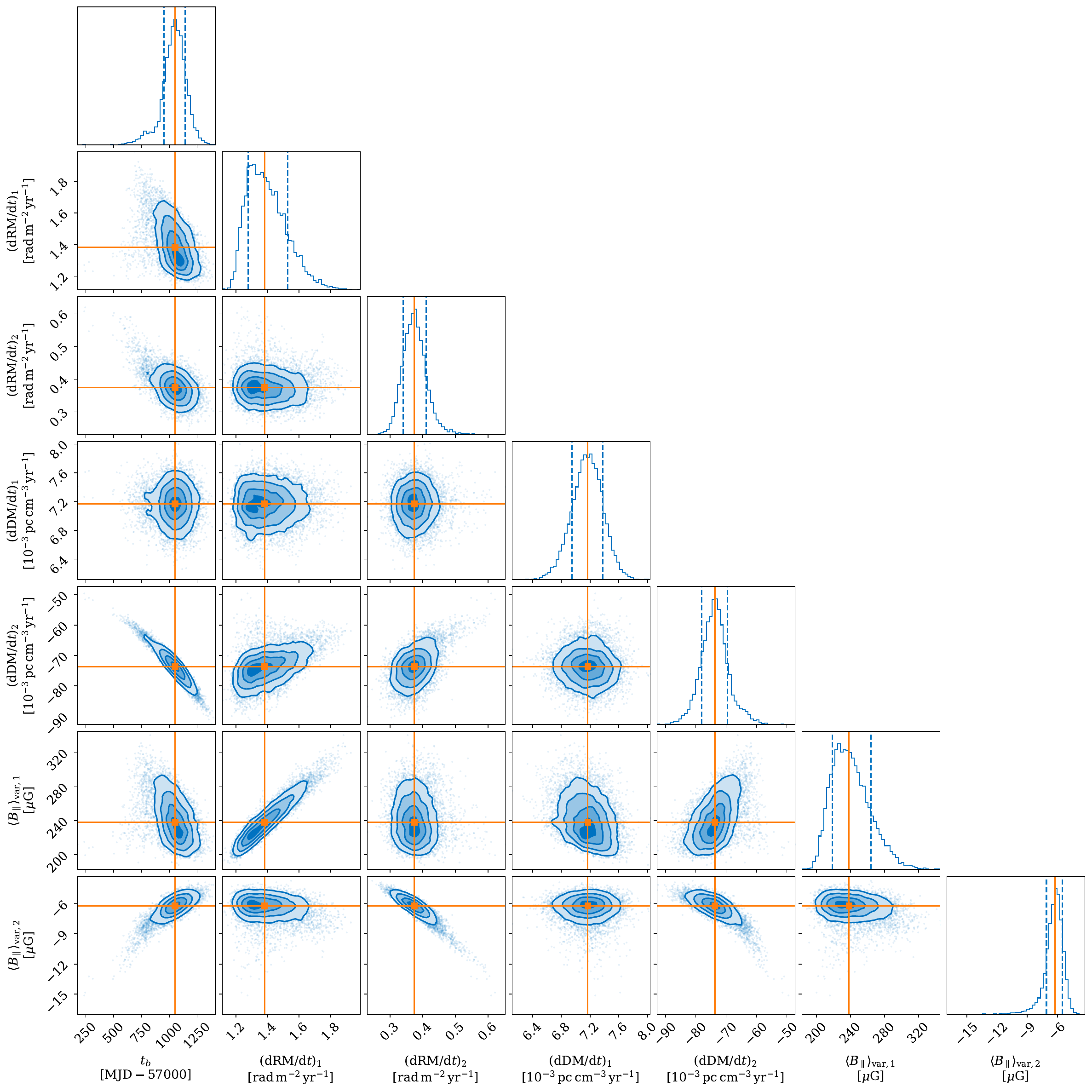}
    \caption{Posterior distributions for the piecewise linear model fit. The orange lines indicate the median and the blue dashed lines indicate the 16\% and 84\% percentiles of 1D posterior distributions. The contours indicate the 11.8\%, 39.3\%, 67.5\%, and 86.4\% percentiles of the 2D posterior distributions.}
    \label{fig:pos_pw}
\end{figure*}

The posterior distributions of the fit parameters and \Bparvar{} are shown in Figures \ref{fig:pos_lin} and \ref{fig:pos_pw}.

\end{document}